\documentclass[twocolumn]{aa}
\usepackage{booktabs}   
\usepackage{arydshln}   
\usepackage{amsmath}
\usepackage{txfonts}
\usepackage{booktabs}
\usepackage{float}
\usepackage{amsmath}
\usepackage[utf8]{inputenc}
\usepackage{graphicx}
\usepackage{float}
\usepackage{hyperref}
\usepackage{csvsimple}
\usepackage{adjustbox}
\usepackage{natbib}

\def\msun{\hbox{M$_\odot$}}

\begin{document}

\title{Astrophysical properties of star clusters projected toward tidally
perturbed SMC regions}

\author{Denis M.F. Illesca\inst{1,2}\thanks{\email{denisillesca1113@gmail.com}}
\and  Andr\'es E. Piatti\inst{1,2} \and  Mat\'{\i}as Chiarpotti\inst{1,2} \and Roberto Butr\'on\inst{1}}

\institute{Instituto Interdisciplinario de Ciencias B\'asicas (ICB), CONICET-UNCUYO, Padre J. Contreras 1300, M5502JMA, Mendoza, Argentina;
\and Consejo Nacional de Investigaciones Cient\'{\i}ficas y T\'ecnicas (CONICET), Godoy Cruz 2290, C1425FQB,  Buenos Aires, Argentina\\
}

\date{Received / Accepted}

\abstract{We report on the astrophysical properties of a sample of star clusters in the Small 
Magellanic Cloud (SMC). They have been selected with the aim of looking for the
connection between their ages, heliocentric distances and metallicities with the existence
of tidally perturbed/induced outermost SMC regions. We derived the star cluster fundamental
parameters from relatively deep Survey of the Magellanic Stellar History
(SMASH) DR2 color-magnitude diagrams (CMDs), cleaned from field star contamination, and
compared to thousand synthetic CMDs covering a wide range of heliocentric distances, ages
and metal content. Heliocentric distances for 15 star clusters are derived for the first
time, which represents an increase of $\sim$50 per cent of SMC clusters with estimated
heliocentric distances. The analysis of the age-metallicity relationships (AMRs) 
of cluster located in outermost regions distributed around the SMC and in the SMC Main Body
reveals that they have followed the overall galaxy chemical enrichment history.
However, since half of the studied clusters are placed in front of or behind the
SMC Main Body, we concluded that they formed in the SMC and have traveled outward 
because of the tidal effects from the interaction with the Large Magellanic Cloud (LMC). 
Furthermore, metal-rich clusters formed recently in some of these outermost regions
from gas that was also dragged by tidal effects from the inner SMC.
This outcome leads to consider the SMC as a galaxy scarred by the LMC tidal interaction 
with distance-perturbed  and newly induced outermost stellar substructures.}


\titlerunning{SMC star clusters}

\authorrunning{D.M.F. Illesca et al.}

\maketitle

\markboth{D.M.F. Illesca et al.: }{SMC star clusters}

\section{Introduction}

The Magellanic Clouds is a pair of interacting irregular dwarf galaxies that 
are experiencing their first passages close to the Milky Way \citep{kallivayalil2013third}. 
Indeed, since recent years, some trails of the tidal effects on them have been observed. 
In the case of the Small Magellanic Cloud (SMC) several efforts have been made with the aim 
of precisely tracing its 
internal kinematic behavior 
\citep[][, and references therein]{niederhofer2018vmc,zivick2018proper,di2019dynamics}. 
However, depending on the stellar population  chosen and the observational data used, such 
tidal perturbations turned out to be of different magnitude 
\citep{de2020revealing,zivick2021deciphering,niederhofer2021vmc}.
As far as we are aware, tidal effects have been found to be stronger in the outermost
galaxy regions, particularly those facing to the other interacting galaxy, while they are
close to each other \citep{almeida2024exploring,cullinane2023magellanic}.
Until now, SMC star clusters have not been extensively used as tracers  of the internal
 kinematics of this galaxy \citep{maia2019viscacha,piatti2021kinematics}. However, they 
present  several advantages from different points of view. For instance, their mean radial
 velocities  and proper motions come from the average of several measurements of cluster
 members, thus  resulting in mean values based on a sound statistics. Star clusters are 
also appropriate representatives of  the internal motion in the SMC, compared to the radial
velocities and mean proper motions of stars aligned along different lines of sight through it. 
This is also valid even when it comes from the separate analysis of different stellar populations of 
the galaxy field. On the other hand, star clusters have precisely determined ages and distances, 
so that the internal kinematics of the SMC can be easily linked to the metallicities 
of the star clusters, and thus provide an adequate framework for our understanding  of their 
formation and chemical and kinematic evolution of the galaxy.

\setlength{\arrayrulewidth}{0.01pt}
\renewcommand{\arraystretch}{1.3}\begin{table*}[htbp]
    \centering
    \caption{Averaged literature values of fundamental parameters of the studied star
clusters.}    
    \label{tab1}
    \begin{tabular}{c c c c c c}
        \toprule
        \toprule
        Star Cluster & $(m-M)_0$ & Log(age /yr) & [Fe/H] & Method & References \\
         & (mag) &  & (dex) &  &  \\
         \midrule
B88 & 19.0 $\pm$ 0.10 & 8.12 $\pm$ 0.22 & -0.25 $\pm$ 0.15  & Photometry & 1;2 \\
B99 & --- & 8.10 $\pm$ 0.01 & -0.84 $\pm$ 0.04 & Photometry; CaII triplet & 5;6 \\
B139 & 19.10 $\pm$ 0.11 & 8.06 $\pm$ 0.12 & -0.30 $\pm$ 0.24 & Photometry & 1;2 \\
B168 & 18.96 $\pm$ 0.07 & 9.82 $\pm$ 0.06 & -1.08 $\pm$ 0.06  & Photometry; CaII triplet & 3;4 \\
BS116 & 18.75 $\pm$ 0.12 & 9.27 $\pm$ 0.07 & -0.77 $\pm$ 0.12  & Photometry & 1 \\
BS121 & --- & 9.45 $\pm$ 0.08 & -0.66 $\pm$ 0.07 & CaII triplet & 7;8 \\
BS188 & 18.61 $\pm$ 0.12 & 9.26 $\pm$ 0.05 & -0.94 $\pm$ 0.06 & Photometry; CaII triplet & 3;4 \\
H86-97 & --- & 8.50 $\pm$ 0.02 & -0.70 $\pm$ 0.04 & Photometry; CaII triplet & 6;9 \\
HW31 & --- & 9.28 $\pm$ 0.11 & -0.89 $\pm$ 0.04 & CaII triplet; Photometry & 2;8;10 \\
HW41 & --- & 9.38 $\pm$ 0.29 & -0.67 $\pm$ 0.05  & CaII triplet; Photometry & 2;8;10 \\
HW47 & --- & 9.48 $\pm$ 0.07 & -0.96 $\pm$ 0.22  & CaII triplet; Photometry & 8;11;12 \\
HW56 & 18.69 $\pm$ 0.09 & 9.47 $\pm$ 0.04 & -0.90 $\pm$ 0.14 & Photometry; CaII triplet & 1;4 \\
HW64 & 19.19 $\pm$ 0.08 & 7.47 $\pm$ 0.31 & -0.38 $\pm$ 0.20 & Photometry & 1 \\
HW67 & 18.71 $\pm$ 0.13 & 9.35 $\pm$ 0.05 & -0.74 $\pm$ 0.12 & Photometry; CaII triplet & 1;6 \\
HW73 & 19.18 $\pm$ 0.04 & 8.01 $\pm$ 0.08 & -0.32 $\pm$ 0.22 & Photometry & 1;2 \\
HW77 & 18.83 $\pm$ 0.12 & 9.05 $\pm$ 0.04 & -1.02 $\pm$ 0.11 & Photometry & 13 \\
HW84 & --- & 9.29 $\pm$ 0.05 & -0.90 $\pm$ 0.05 & Photometry & 7;8;12 \\
HW86 & 18.53 $\pm$ 0.09 & 9.15 $\pm$ 0.05 & -0.65 $\pm$ 0.09 & CaII triplet; Photometry & 7;8;14 \\
IC1655 & 18.66 $\pm$ 0.04 & 8.18 $\pm$ 0.20 & -0.61 $\pm$ 0.18 & Photometry & 1;2 \\
L2 & 18.72 $\pm$ 0.10 & 9.59 $\pm$ 0.02 & -1.29 $\pm$ 0.13 & Photometry; CaT lines & 13;21;22 \\
L3 & 18.64 $\pm$ 0.14 & 9.08 $\pm$ 0.11 & -0.75 $\pm$ 0.33 & Photometry & 10;21;24 \\
L4 & 18.65 $\pm$ 0.10 & 9.90 $\pm$ 0.06 & -1.08 $\pm$ 0.04 & Photometry; CaII triplet & 3;7;8 \\
L6 & 18.70 $\pm$ 0.09 & 9.94 $\pm$ 0.06 & -1.24 $\pm$ 0.03 & Photometry; CaII triplet & 3;7;8 \\
L7 & --- & 9.17 $\pm$ 0.04 & -0.82 $\pm$ 0.06 & CaII triplet; Photometry & 6;7;20 \\
L9 & 18.93 $\pm$ 0.07 & 9.22 $\pm$ 0.07 & -0.66 $\pm$ 0.01 & Photometry; CaII triplet & 6;12;13 \\
L11 & --- & 9.48 $\pm$ 0.06 & -1.07 $\pm$ 0.18 & Deep Photometry; CaII triplet & 7;8;15 \\
L13 & 18.66 $\pm$ 0.04 & 8.75 $\pm$ 0.08 & -1.05 $\pm$ 0.06 & Photometry; CaII triplet & 6;13;18 \\
L17 & --- & 9.64 $\pm$ 0.06 & -0.84 $\pm$ 0.03 & CaII triplet & 7;8 \\
L19 & --- & 9.48 $\pm$ 0.03 & -0.87 $\pm$ 0.02 & Photometry; CaII triplet & 1;19;20 \\
L27 & 18.84 $\pm$ 0.10 & 9.55 $\pm$ 0.06 & -1.01 $\pm$ 0.10 & Photometry; CaII triplet & 2;3;7;19;22;23 \\
L58 & --- & 9.13 $\pm$ 0.28 & -0.78 $\pm$ 0.02 & Photometry; CaII triplet & 2;6;25 \\
L73 & 18.83 $\pm$ 0.13 & 9.27 $\pm$ 0.04 & -0.65 $\pm$ 0.13 & Photometry & 1 \\
L95 & 19.18 $\pm$ 0.10 & 7.94 $\pm$ 0.34 & -0.36 $\pm$ 0.27 & Photometry & 1;2 \\
L100 & 18.81 $\pm$ 0.06 & 9.46 $\pm$ 0.04 & -0.75 $\pm$ 0.08 & Photometry; CaII triplet & 1;4;11 \\
L108 & --- & 9.27 $\pm$ 0.09 & -0.88 $\pm$ 0.18 & CaII triplet; Photometry & 7;8 \\
L110 & 18.95 $\pm$ 0.10 & 9.88 $\pm$ 0.06 & -1.03 $\pm$ 0.05 & Photometry; CaII triplet & 14;7;8 \\
L116 & --- & 9.44 $\pm$ 0.20 & -1.0 $\pm$ 0.11 & Photometry; CaII triplet & 16;17 \\
NGC416 & 18.96 $\pm$ 0.07 & 9.78 $\pm$ 0.04 & -0.85 $\pm$ 0.04 & Photometry; CaII triplet & 23;21;26 \\
NGC458 & 18.74 $\pm$ 0.03 & 8.14 $\pm$ 0.07 & -0.39 $\pm$ 0.08 & Photometry & 1;27 \\
OGLE-SMC 133 & --- & 9.80 $\pm$ 0.40 & -0.80 $\pm$ 0.07 & Photometry; CaII triplet & 2;6 \\
        \bottomrule
    \end{tabular}
   \tablebib{
(1) \citet{piatti2022revisiting}; 
(2) \citet{glatt2010ages}; 
(3) \citet{piatti2022revisiting}; 
(4) \citet{dias2021viscacha}; 
(5) \citet{maia2014mass}; 
(6) \citet{parisi2015ii}; 
(7) \citet{parisi2009ii}; 
(8) \citet{parisi2014age}; 
(9) \citet{chiosi2006age}; 
(10) \citet{de2022ii}; 
(11) \citet{parisi2022ii}; 
(12) \citet{piatti2005tracing}; 
(13) \citet{saroon2023viscacha}; 
(14) \citet{oliveira2023viscacha}; 
(15) \citet{livanou2013age}; 
(16) \citet{piatti2001ages}; 
(17) \citet{parisi2022ii}; 
(18) \citet{nayak2018star}; 
(19) \citet{narloch2021metallicities}; 
(20) \citet{gatto2021step}; 
(21) \citet{de2015clustering}; 
(22) \citet{dias2022viscacha}; 
(23) \citet{milone2023hubble}; 
(24) \citet{dias2014self}; 
(25) \citet{maia2019viscacha}; 
(26) \citet{glatt2008age}; 
(27) \citet{alcaino2003small}.
}
\end{table*}

\begin{figure}
\includegraphics[width=\columnwidth]{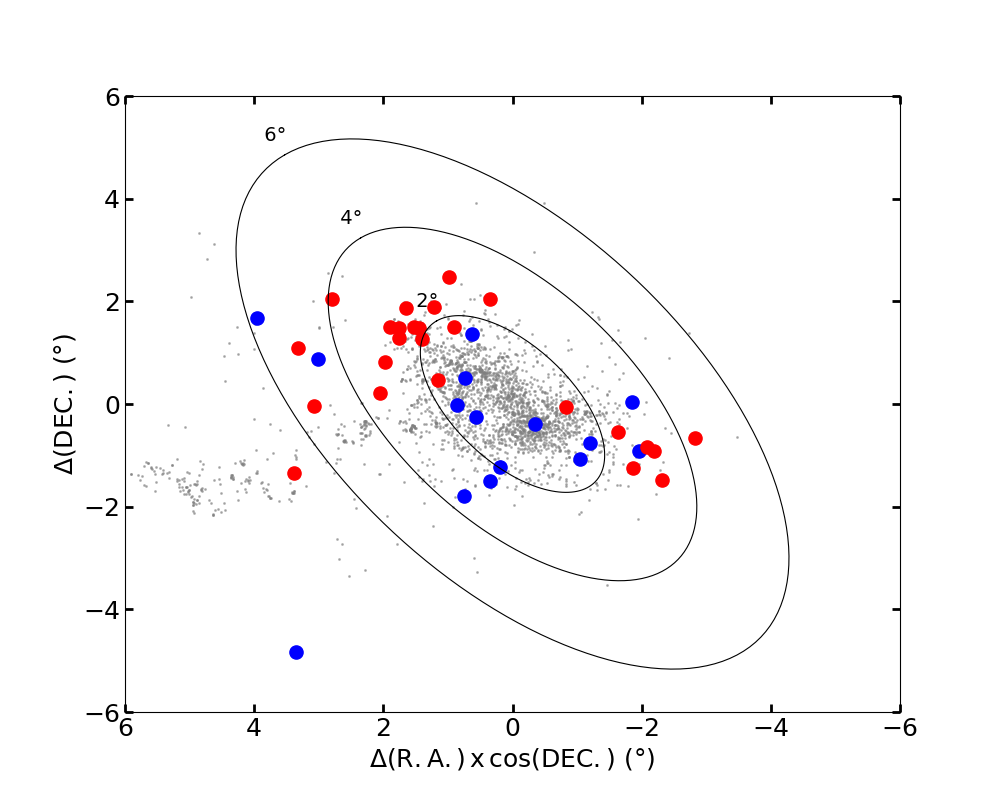}
\caption{Spatial distribution of SMC clusters \citep{bicaetal2020} represented
with gray points. Red and blue filled circles represent the studied clusters with
or without previous estimates of their heliocentric distances, respectively.
The ellipses correspond to the framework devised by \cite{piatti2007five} for
$a$= 2$\degr$, 4$\degr$, and 6$\degr$, respectively.}
\label{fig1}
\end{figure}

By studying SMC star clusters distributed across the Southern Bridge, the SMC Wing, the
West Halo, the Counter Bridge \citep{dias2014self}, among other tidally perturbed/induced SMC 
stellar substructures,  it is possible to shed light on the origin of such stellar 
substructures, on their dimensions and their location with respect to the SMC Main Body, 
their chemistry and their epoch of formation \citep[see, e.g.][]{piatti2022revisiting}. Precisely,
we have embarked in an effort of homogeneously estimating astrophysical properties (e.g., heliocentric 
distances, ages, and metallicities) of star clusters distributed in those SMC selected regions, 
aiming at performing an overall analysis of the magnitude of the tidal effects throughout the 
whole galaxy. We particularly focus on the estimation, for the first time, of star cluster heliocentric 
distances, which are remarkably lacking in the literature \citep{piatti2023depth}. During  several 
decades, SMC star clusters were thought to be located at the mean galaxy distance, 
with the consequent misleading in the interpretation of their formation history in the galaxy.
However, several studies have shown that the SMC is much more extended along the line of sight 
than its apparent size projected in the sky \citep{ripepi2017vmc,graczyk2020distance}, which
makes meaningful to derive accurate heliocentric distances.

This work is organized as follows: Section 2 describes the selection of the star cluster sample, 
alongside with the data used in the analysis, the assignment of cluster star membership, and 
the estimation of the cluster astrophysics parameters, namely, reddening,
age, distance and metallicity. Section 3 deals with the resulting clusters' properties, while 
in Section 4 we analyze them to the light of the SMC stellar spatial distribution and star 
formation history. Section 5 summarizes the main conclusions of this work.

\section{Observational data}

We selected 40 star clusters mainly distributed across the outer SMC regions
using the catalog compiled by \cite{bicaetal2020}. Table ~\ref{tab1}  gathers the most
relevant fundamental parameters available in the literature, alongside with the method
employed to derive them. We used as reference the framework devised by \cite{piatti2007five}, 
which consists of concentric equally aligned ellipses,
with a position angle of 54$\degr$ (measured anti-clockwise from the North) and a ratio 
between the semi-major to the semi-minor axes of 2. The selected star clusters are preferentially 
located outside the area covered by an ellipse with a semi-major axis of 2$\degr$. As 
mentioned in Section~1, they populate known perturbed/induced stellar substructures.
Figure~\ref{fig1} illustrates the spatial distribution of the selected cluster sample, where
fifteen clusters that have not been targeted for estimating their heliocentric distances 
until now are highlighted with blue symbols. 

\begin{figure}
\includegraphics[width=\columnwidth]{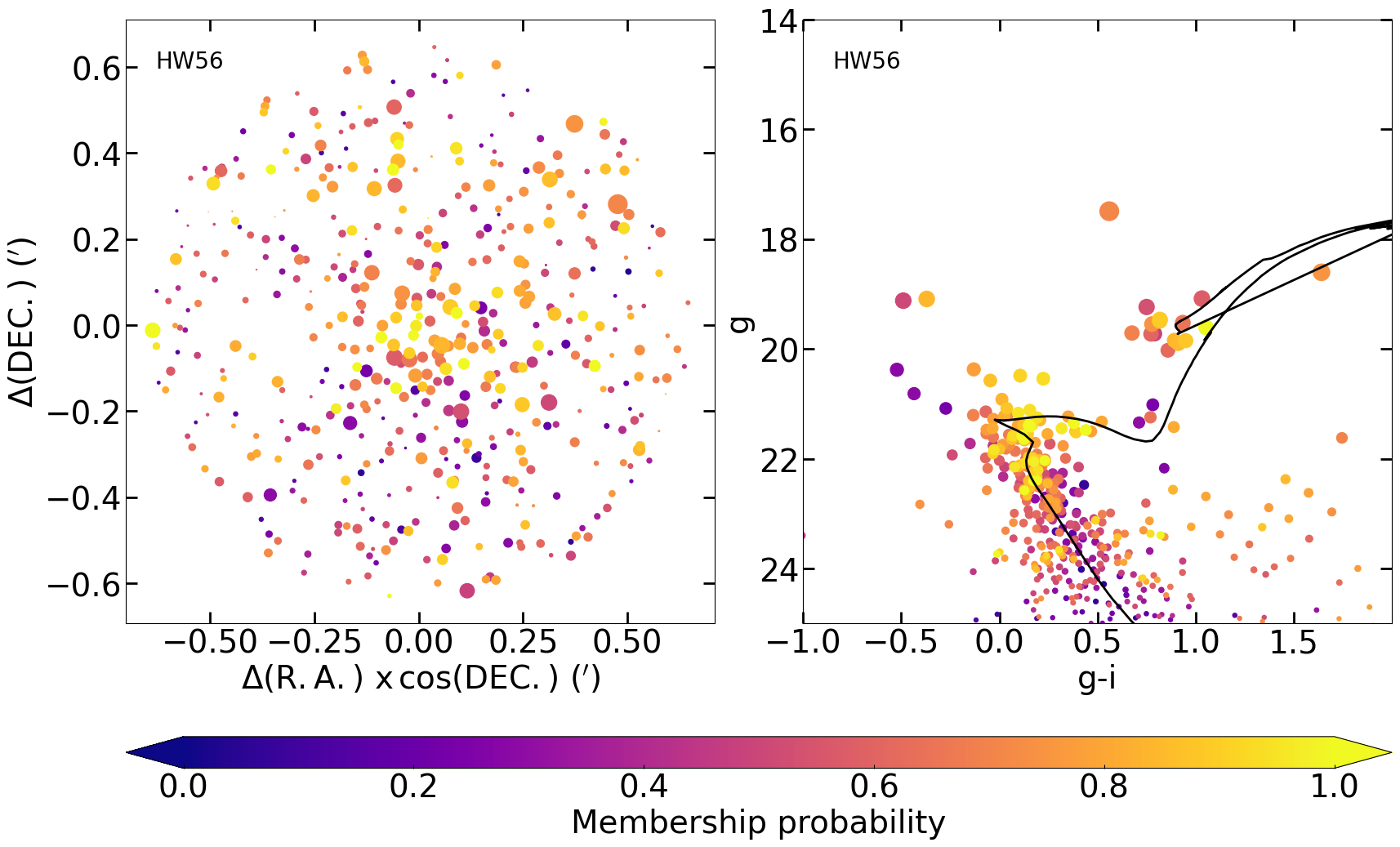}
\caption{{\bf Left:} schematic finding chart of HW~56. The size of the symbols
are proportional to the star brightness. {\bf Right:} Cluster color-magnitude diagram
with the theoretical isochrone \citep{bressan2012parsec} corresponding to the
adopted cluster parameters superimposed (see Table~\ref{tab2}). The color bar
indicates the membership probability.}
\label{fig2}
\end{figure}

Considering the overall low brightness and compactness of the selected star 
clusters, the relatively deep photometric data provided by the Data Release 2 of the
Survey of the Magellanic Stellar History \cite[SMASH, ][]{nideveretal2021} is suitable to 
perform an homogeneous estimation of their astrophysical parameters.
Its limiting magnitude is beyond the 24th magnitude in the outskirts of the 
SMC, which is several magnitudes underneath the Main Sequence turnoff of a
$\sim$ 1 Gyr old cluster located at the distance of the SMC. The data of each selected cluster were 
retrieved from the Astro Data Lab\footnote{https://datalab.noirlab.edu/smash/smash.php}
interface, 
which is part of the Community Science and Data Center at NSF’s National Optical Infrared 
Astronomy Research Laboratory. Particularly, we were interested in the R.A and 
Dec. coordinates, the PSF $g,i$ magnitudes and their respective errors, the $E(B-V)$ 
interstellar reddening and $\chi$ and {\sc sharpness} parameters of stellar sources
located inside a circular region with a radius six times that of the cluster \citep{bicaetal2020}. 
With the aim of minimizing the contamination of bad pixels, cosmic 
rays, background galaxies, and unrecognized double stars, we selected those sources 
with 0.2 $\le$ {\sc sharpness} $\le$ 1.0 and $\chi$ $<$ 0.5. 

Aiming at disentangling the main features of the observed cluster color-magnitude 
diagrams (CMDs), we first cleaned them from star field contamination on a statistical basis
using the procedure devised by \cite{piatti2012washington}. It uses the magnitude and color of 
each star in the so-called reference star field CMD and finds the closest one in magnitude 
and color in the cluster's CMD and subtracts it. We used the reddening corrected $g_0$,$i_0$
magnitudes computed from the observed $g,i$ magnitudes, the $E(B-V)$ values retrieved 
from SMASH and the $A_\lambda$/$E(B-V)$ ratios, for $\lambda$ = $g,i$, given by  Abott et al. 
(2018). In practice, (1) we selected stars located within the cluster circle with a radius
equal the cluster's radius \citep{bica2020vizier}, and built the respective CMD; (2) we constructed 
a thousand CMDs with an area equal to the cluster circle and randomly located at a distance from 
the cluster center equals to 4.5 times the cluster's radius; (3) we performed the subtraction from 
the cluster's CMD of stars distributed similarly as those in the
reference field CMDs, employing in the decontamination process one reference field CMD
at a time. Therefore, we generated a thousand different cleaned cluster's CMDs. (4) 
Finally, we assigned membership probabilities based on the number of times a star
survived the cleaning procedure, namely: $P = 1-N/1000$, where $N$ represents the number
of times a star was subtracted. Figure ~\ref{fig2} illustrates 
the resulting membership probability distribution of stars in the field of HW56.

From the field star decontaminated cluster photometry we derived the clusters' properties
employing routines of the Automated Stellar Cluster Analysis code \citep[ASteCA,][]{perren2015asteca}. 
ASteCA allows to derive simultaneously the metallicity, the age, the distance, and the cluster present 
mass by producing a large number of synthetic CMDs and then choosing that that 
best reproduces the cleaned star cluster CMD. The metallicity, age, distance, cluster present mass 
and binary fraction associated to that generated synthetic CMD are adopted as the best-fitted 
cluster properties. In order to generate a statistical significant sample of synthetic
CMDs, we adopted the initial mass function of \cite{kroupa2002initial} and a minimum mass ratio for the 
generation of binaries of 0.5. Cluster masses and binary 
fractions were set in the ranges 100-5000 $\msun$ and 0.0-0.5, respectively. 
We also used PARSEC\footnote{http://stev.oapd.inaf.it/cgi-bin/cmd} v1.2S isochrones 
\citep{bressan2012parsec} for the SMASH photometric system, spanning metallicities 
$Z$ from 0.0001 ([Fe/H] = -2.18 dex) up dex up to 0.030 ([Fe/H] = 0.30 dex), 
in steps of $\Delta$$Z$ = 0.001, and log(age /yr) from 6.0 (1 Myr) up to 10.1 (12.5 Gyr) in 
steps $\Delta$(log(age /yr)) = 0.025. These metalllcity and age ranges embrace the
whole SMC age-metallicity relationship \citep{pg13}.In some cases, we used
smaller range values as prior to secure astrophysically meaningful results.
We entered ASteCA with the cluster SMASH photometry 
and the previously independently assigned membership probabilities to each observed 
star and derived the aforementioned cluster astrophysical parameters with their
respective uncertainties. These errors were estimated from the standard bootstrap method 
described in \cite{efron1982jackknife}. Table~\ref{tab2} lists the resulting cluster parameters, 
while Figure~\ref{fig2} depicts the CMD of HW56 with the isochrone corresponding to the clusters'
parameters superimposed.
Similar figures for the remaining cluster sample are included in the Appendix.

\begin{figure}
\includegraphics[width=\columnwidth]{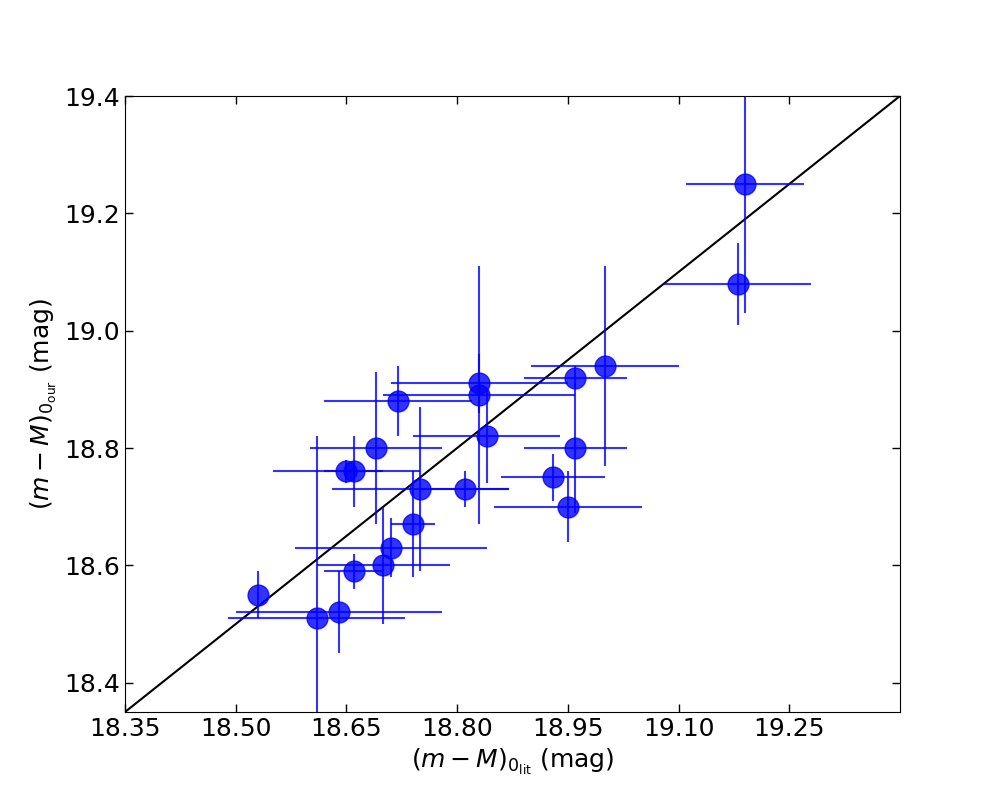}
\caption{Comparison between the derived distances moduli and those adopted from the literature.
The solid line represents the identity relationship.}
\label{fig3}
\end{figure}

\begin{figure}
\includegraphics[width=\columnwidth]{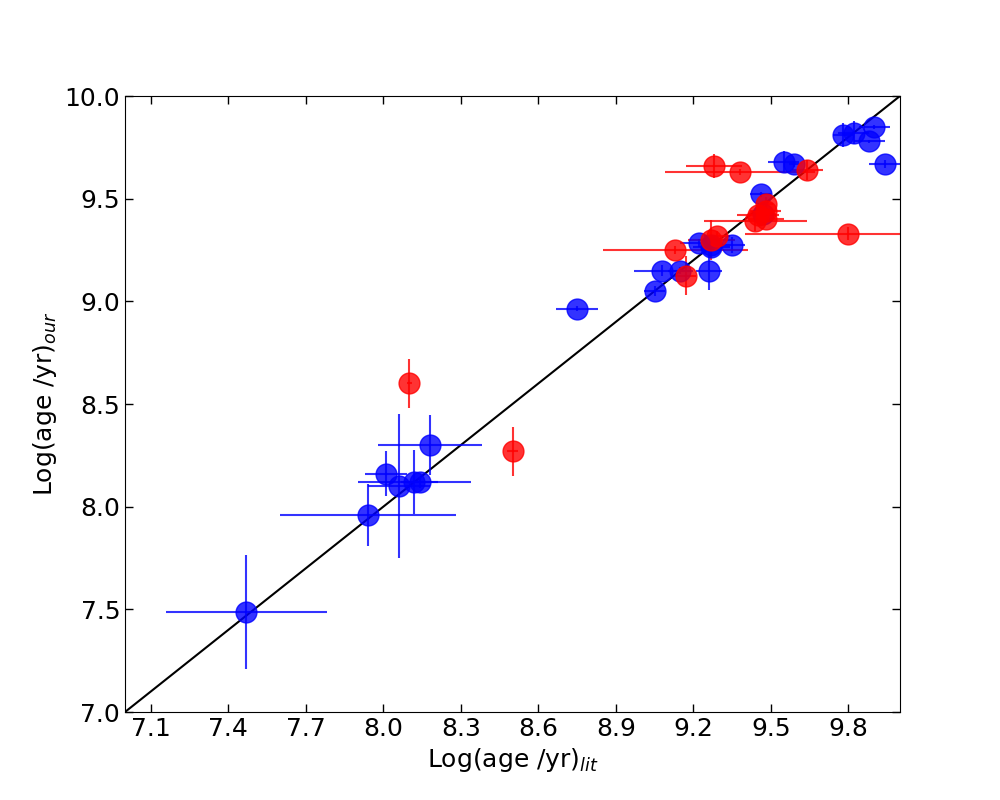}
\caption{Comparison between the derived ages and those adopted from the literature. Blue filled circles represent star clusters with previous estimates of their heliocentric distances, while red filled circles correspond to those without such estimates. The solid line represents the identity relation.}
\label{fig4}
\end{figure}

\begin{figure}
\includegraphics[width=\columnwidth]{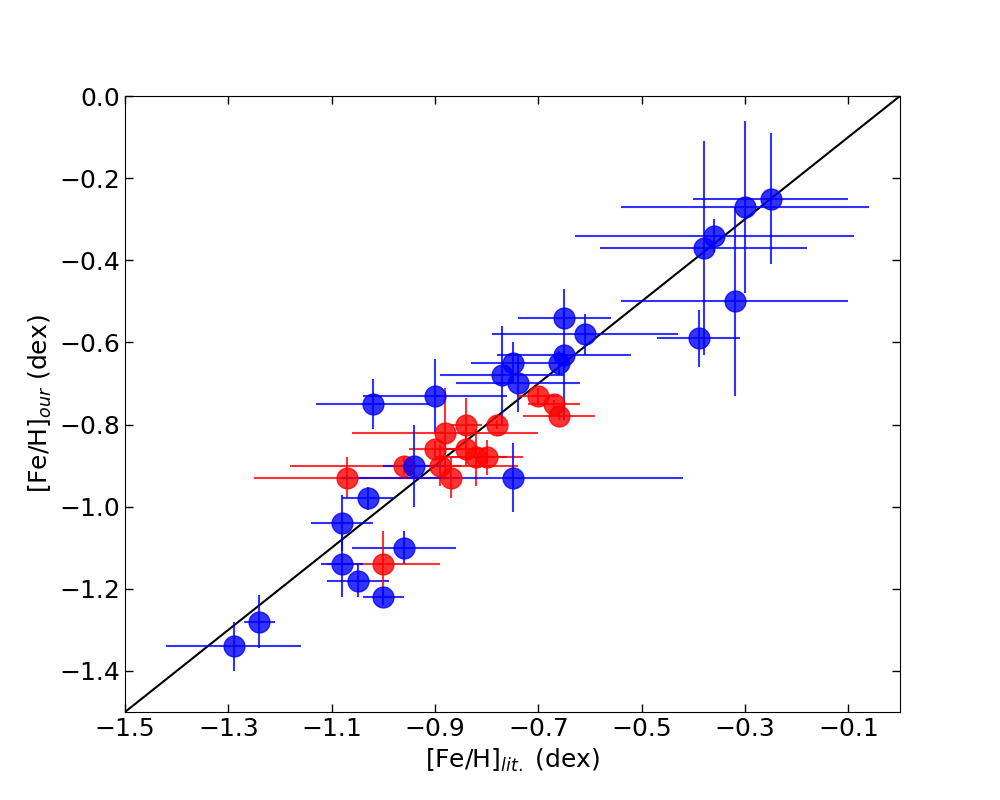}
\caption{Comparison between the derived metallicities and those adopted from the literature. Blue filled circles represent star clusters with previous estimates of their heliocentric distances, while red filled circles correspond to those without such estimates. The solid line represents the identity relation.}
\label{fig5}
\end{figure}

\setlength{\arrayrulewidth}{0.03pt}
\renewcommand{\arraystretch}{1.3}
\begin{table*}
    \caption{Derived fundamental parameters of the studied star
clusters.} 
\label{tab2}   
\begin{tabular}{c c c c c c }\hline\hline
Star Cluster & $(m-M)_0$  & $d$ & [Fe/H] & Log(age /yr) & SMC region$^a$ \\
 & (mag) & (kpc) & (dex) &  &  \\\hline
B88 & 18.94 $\pm$ 0.17 & 61.38 $\pm$ 4.74 & -0.25 $\pm$ 0.16 & 8.12 $\pm$ 0.16 & Counter Bridge \\
B99 & 17.90 $\pm$ 0.08 & 38.02 $\pm$ 1.38 & -0.86 $\pm$ 0.04 & 8.6 $\pm$ 0.12 & Main Body \\
B139 & 18.83 $\pm$ 0.26 & 58.34 $\pm$ 6.89 & -0.27 $\pm$ 0.21 & 8.10 $\pm$ 0.35 & Northern Bridge \\
B168 & 18.80 $\pm$ 0.11 & 57.54 $\pm$ 2.64 & -1.04 $\pm$ 0.07 & 9.82 $\pm$ 0.06 & Northern Bridge \\
BS116 & 18.73 $\pm$ 0.14 & 55.72 $\pm$ 3.54 & -0.68 $\pm$ 0.12 & 9.26 $\pm$ 0.05 & Main Body \\
BS121 & 19.38 $\pm$ 0.01 & 75.16 $\pm$ 0.34 & -0.78 $\pm$ 0.01 & 9.42 $\pm$ 0.01 & Main Body \\
BS188 & 18.51 $\pm$ 0.31 & 50.35 $\pm$ 7.09 & -0.90 $\pm$ 0.10 & 9.15 $\pm$ 0.10 & Northern Bridge \\
H8697 & 17.83 $\pm$ 0.20 & 36.81 $\pm$ 3.35 & -0.73 $\pm$ 0.02 & 8.27 $\pm$ 0.12 & Main Body \\
HW31 & 18.70 $\pm$ 0.10 & 54.95 $\pm$ 2.50 & -0.90 $\pm$ 0.05 & 9.66 $\pm$ 0.06 & Southern Bridge \\
HW41 & 18.60 $\pm$ 0.05 & 52.48 $\pm$ 1.19 & -0.75 $\pm$ 0.01 & 9.61 $\pm$ 0.02 & Main Body \\
HW47 & 18.70 $\pm$ 0.06 & 54.95 $\pm$ 1.50 & -0.90 $\pm$ 0.02 & 9.40 $\pm$ 0.02 & Southern Bridge \\
HW56 & 18.80 $\pm$ 0.13 & 57.54 $\pm$ 3.40 & -0.73 $\pm$ 0.09 & 9.46 $\pm$ 0.05 & Northern Bridge \\
HW64 & 19.25 $\pm$ 0.22 & 70.79 $\pm$ 7.08 & -0.37 $\pm$ 0.26 & 7.49 $\pm$ 0.28 & Northern Bridge \\
HW67 & 18.63 $\pm$ 0.05 & 53.21 $\pm$ 1.21 & -0.70 $\pm$ 0.07 & 9.28 $\pm$ 0.04 & Northern Bridge \\
HW73 & 18.79 $\pm$ 0.19 & 57.28 $\pm$ 4.95 & -0.50 $\pm$ 0.23 & 8.16 $\pm$ 0.11 & Northern Bridge \\
HW77 & 18.91 $\pm$ 0.05 & 60.53 $\pm$ 1.38 & -0.75 $\pm$ 0.06 & 9.05 $\pm$ 0.02 & Bridge \\
HW84 & 18.30 $\pm$ 0.08 & 45.71 $\pm$ 1.66 & -0.86 $\pm$ 0.02 & 9.32 $\pm$ 0.02 & Northern Bridge \\
HW86 & 18.55 $\pm$ 0.04 & 51.29 $\pm$ 0.93 & -0.54 $\pm$ 0.02 & 9.15 $\pm$ 0.02 & Bridge \\
IC1655 & 18.76 $\pm$ 0.06 & 56.49 $\pm$ 1.54 & -0.58 $\pm$ 0.05 & 8.30 $\pm$ 0.14 & Northern Bridge \\
L2 & 18.88 $\pm$ 0.06 & 59.70 $\pm$ 1.63 & -1.34 $\pm$ 0.06 & 9.67 $\pm$ 0.02 & West Halo \\
L3 & 18.52 $\pm$ 0.07 & 50.58 $\pm$ 1.61 & -0.93 $\pm$ 0.08 & 9.15 $\pm$ 0.03 & West Halo \\
L4 & 18.76 $\pm$ 0.02 & 56.49 $\pm$ 0.51 & -1.14 $\pm$ 0.08 & 9.85 $\pm$ 0.01 & West Halo \\
L6 & 18.60 $\pm$ 0.10 & 52.48 $\pm$ 2.39 & -1.28 $\pm$ 0.07 & 9.67 $\pm$ 0.02 & West Halo \\
L7 & 18.76 $\pm$ 0.09 & 56.49 $\pm$ 2.31 & -0.88 $\pm$ 0.07 & 9.13 $\pm$ 0.09 & West Halo \\
L9 & 18.75 $\pm$ 0.04 & 56.23 $\pm$ 1.02 & -0.65 $\pm$ 0.03 & 9.29 $\pm$ 0.02 & West Halo \\
L11 & 18.87 $\pm$ 0.07 & 59.43 $\pm$ 1.89 & -0.93 $\pm$ 0.05 & 9.44 $\pm$ 0.01 & West Halo \\
L13 & 18.59 $\pm$ 0.03 & 52.24 $\pm$ 0.71 & -1.18 $\pm$ 0.04 & 8.97 $\pm$ 0.01 & West Halo \\
L17 & 18.60 $\pm$ 0.21 & 52.48 $\pm$ 5.01 & -0.80 $\pm$ 0.07 & 9.64 $\pm$ 0.05 & Main Body \\
L19 & 18.75 $\pm$ 0.08 & 56.23 $\pm$ 2.04 & -0.93 $\pm$ 0.05 & 9.48 $\pm$ 0.03 & Main Body \\
L27 & 18.82 $\pm$ 0.08 & 58.07 $\pm$ 1.81 & -1.10 $\pm$ 0.04 & 9.68 $\pm$ 0.05 & Main Body \\
L58 & 18.64 $\pm$ 0.04 & 53.46 $\pm$ 0.97 & -0.80 $\pm$ 0.01 & 9.25 $\pm$ 0.02 & Southern Bridge \\
L73 & 18.89 $\pm$ 0.22 & 59.98 $\pm$ 6.01 & -0.63 $\pm$ 0.16 & 9.27 $\pm$ 0.03 & Counter Bridge \\
L95 & 19.08 $\pm$ 0.07 & 65.46 $\pm$ 2.08 & -0.34 $\pm$ 0.04 & 7.96 $\pm$ 0.15 & Northern Bridge \\
L100 & 18.73 $\pm$ 0.03 & 55.72 $\pm$ 0.76 & -0.65 $\pm$ 0.05 & 9.52 $\pm$ 0.01 & Northern Bridge \\
L108 & 18.67 $\pm$ 0.11 & 54.20 $\pm$ 2.71 & -0.82 $\pm$ 0.11 & 9.30 $\pm$ 0.10 & Northern Bridge \\
L110 & 18.70 $\pm$ 0.06 & 54.95 $\pm$ 1.50 & -0.98 $\pm$ 0.03 & 9.78 $\pm$ 0.01 & Bridge \\
L116 & 18.13 $\pm$ 0.17 & 42.27 $\pm$ 3.27 & -1.14 $\pm$ 0.08 & 9.39 $\pm$ 0.05 & Southern Bridge \\
NGC416 & 18.92 $\pm$ 0.02 & 60.81 $\pm$ 0.46 & -1.22 $\pm$ 0.02 & 9.81 $\pm$ 0.06 & Main Body \\
NGC458 & 18.67 $\pm$ 0.09 & 54.20 $\pm$ 2.22 & -0.59 $\pm$ 0.07 & 8.12 $\pm$ 0.027 & Northern Bridge \\
OGLE-CL SMC 133 & 18.54 $\pm$ 0.04 & 51.05 $\pm$ 0.93 & -0.88 $\pm$ 0.04 & 9.33 $\pm$ 0.03 & Main Body \\\hline
    \end{tabular}

\noindent $^a$ From \citet{dias2014self} and \citet{dias2016smc}.
\end{table*}

\section{Star cluster properties}

Table~\ref{tab2}  shows that the selected cluster sample comprises SMC 
clusters that span a wide age range ($\sim$ 3 Myr - 7.06 Gyr); nearly any known 
SMC metallicity value ([Fe/H] $\sim$  -1.3 dex - -0.25 dex); and a noticeable range 
of heliocentric distances ($d$ $\sim$ 36 kpc - 75 kpc). 
Although some few clusters 
are known to have heliocentric distances that place them outside the Main Body of 
the SMC \citep[$\sim$ 56-62 kpc, see][]{piatti2022revisiting}, we here found nearly 
10 additional clusters also located in the SMC periphery. Because of the impact of 
these new heliocentric distances in our knowledge of the formation and evolution 
of the SMC, we first validated our resulting astrophysical parameters by comparing them 
with those available in the literature (see Table~\ref{tab1}).

Previous estimates of metallicity, age and heliocentric distance of the
studied cluster sample were thoroughly sought in the available literature. 
For clusters with more than one estimate, we averaged the published values,
giving more weight to spectroscopic metallicities than those obtained from
theoretical isochrone fitting to the clusters' CMDs. Whenever ages or heliocentric
distances derived from CMD analyses are available, those from deeper and higher
quality photometry were assigned more weight. For completeness purposes,
we also included fundamental parameters values found in the literature that we 
consider not as accurate as our present ones (B139, HW73).
Table~\ref{tab1} lists the adopted
average values. We here estimated accurate heliocentric 
distances, for the first time, for nearly 40 per cent of the studied clusters.
This outcome points to the need of an observational campaign
to perform a deep imaging survey to definitively tackle the 3D spatial distribution of the
SMC cluster population from accurate heliocentric distance for a 
statistical significant SMC cluster sample. At present less than 5 per cent of the
cataloged SMC star cluster population has some individual heliocentric distance estimate 
\citep{piatti2023depth}. As mentioned above, until recently, photometric 
studies of SMC clusters assumed the SMC mean distance
modulus. Such an assumption was valid in the context of a poorer photometric data, 
which did not distinguish distance variation of
$\sim$ 4.7 kpc ($\Delta$(distance modulus) $\sim$ 0.16 mag); the FWHM of the
SMC Main Body \citep{graczyk2020distance}.

Figures~\ref{fig3} to \ref{fig5} show a comparison between our resulting true distance
moduli, ages and metallicities with those found in the literature, respectively.
We did not included in Figure~\ref{fig3} B139 and HW~73 because the available distance moduli are not as accurate as our present estimates
(this can be checked by visually comparing Figure 2 of 
\citet{piatti2022revisiting}  
with the corresponding ones in the Appendix.
As can be seen, all three parameters show a very good agreement, the difference being
$\Delta$(log(age /yr)$_{our}$ - log(age /yr)$_{lit.}$ = 0.09 $\pm$ 0.07,
$\Delta$([Fe/H]$_{our}$ - [Fe/H]$_{lit.}$ = 0.073 $\pm$ 0.051 dex, and
$\Delta$($(m-M)_0$$_{our}$ - $(m-M)_0$$_{lit.}$) = 0.14 $\pm$ 0.06 mag, respectively.

\subsection{Some selected star clusters}

In this section we focus on the most prominent star clusters included in our sample
without prior estimates of their heliocentric distance, with the aim of describing our resulting parameters in some detail.

For B99 we derive a mean true distance modulus of 17.9 mag, which in turn implies a
mean heliocentric distance $d$ = 38.02 kpc; mean cluster age and metallicity resulted to
be 0.39 Gyr and [Fe/H] = -0.86, respectively. Our derived age is slightly larger than
that obtained by \cite{maia2014mass} (0.12 Gyr), while we found an excellent 
agreement with the spectroscopic metallicity value derived by \cite{parisi2015ii}
([Fe/H] = -0.84 dex).

The best-fitted theoretical isochrone for BS121 corresponds to an age of 2.63 Gyr and 
an overall metal content [Fe/H] = -0.78 dex, for a true distance modulus of
19.38 mag ($d$ = 75.16 kpc). Both estimated cluster age and metallicity are in
very good agreement with the values obtained by  \cite{parisi2009ii} and \cite{parisi2015ii}.

\begin{figure}
\includegraphics[width=\columnwidth]{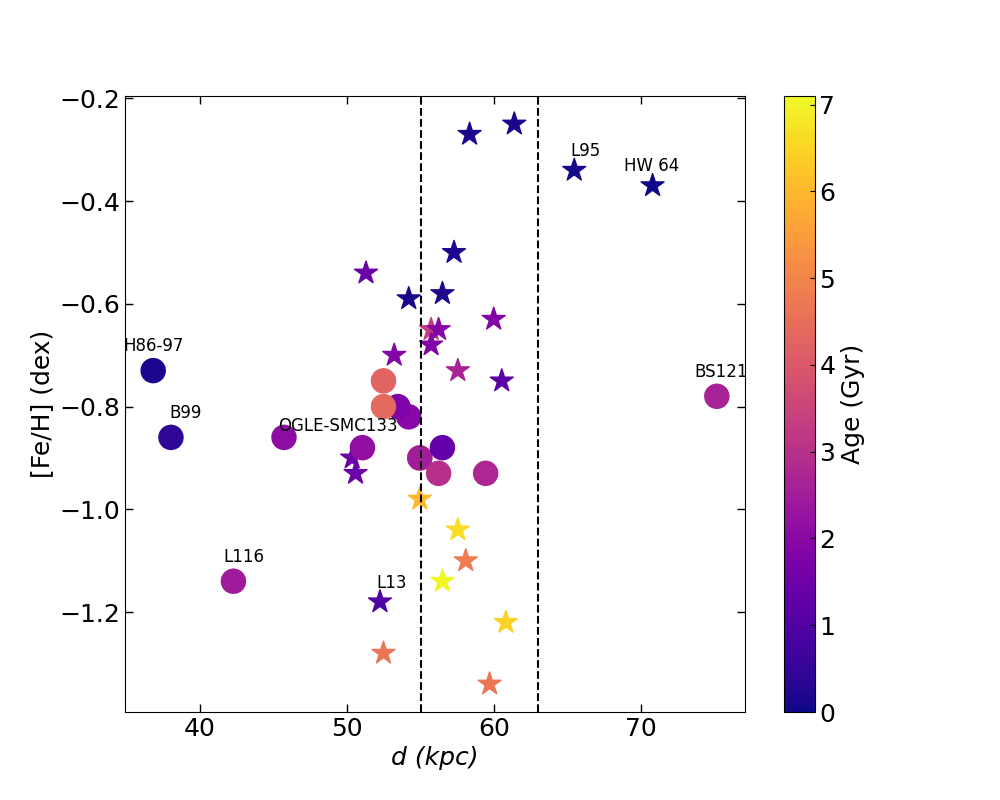}
\caption{Relationship between the derived metallicities and heliocentric distances of the 
studied star clusters, colored according to
their derived ages. Filled stars and circles represent star clusters with or without
previous estimates of their heliocentric distances, respectively. The vertical dashed 
lines delimit the extension of the SMC Main Body \citep{piatti2023depth}.}
\label{fig6}
\end{figure}

The CMD of H86-97 was satisfactorily reproduced by an isochrones of age = 0.19 Gyr and
[Fe/H] = -0.73 dex, for a mean true distance modulus of 17.83 mag ($d$ = 36.81 kpc).
Our estimated age resulted to be slightly younger than the age derived by
\cite{chiosi2006age} (0.32 Gyr), while the present cluster metal content agrees well
with the Ca~II triplet metallicity obtained by  \cite{parisi2015ii}.
\begin{figure}
\includegraphics[width=\columnwidth]{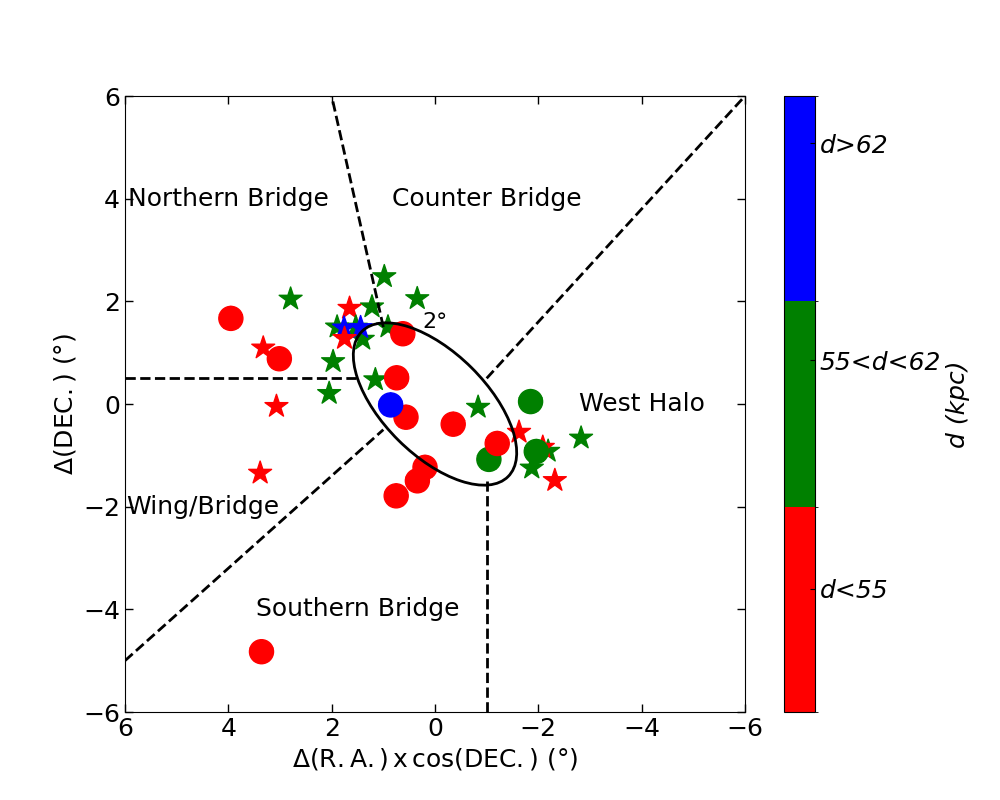}
\caption{Distribution in the sky of the studied SMC star clusters. Colored symbols 
are according to the star cluster heliocentric distances. Dashed lines delimit
the different outermost SMC regions (see text for details).
Filled stars and circles represent star clusters with or without
previous estimates of their heliocentric distances, respectively.}
\label{fig7}
\end{figure}

The isochrone which best reproduces the CMD of OGLE-SMC 133 is that of an age of 2.14 Gyr,
[Fe/H] = -0.88 dex and true distance modulus of 18.54 mag (51.05 kpc). Although the
resulting metallicity is in agreement with that of  \cite{parisi2015ii} ([Fe/H] = -0.80 dex),
our cluster age resulted slightly younger than the value obtained by \cite{glatt2010ages}
(6.29 Gyr).

The ages and metallicities of L7, L19, L58 and L116 are in very good agreement
with the values retrieved from the literature (see Tables~\ref{tab1} and \ref{tab2}), while
the derived mean true distance moduli are 18.76 mag ($d$ = 56.49 kpc), 18.75 mag
($d$ = 56.23 kpc),  18.64 mag ($d$ = 53.46 kpc), and  18.13 mag ($d$ = 42.27 kpc),
respectively.

\section{Analysis and discussion}

In the subsequent analysis we discuss the relationship between the clusters' ages and
metallicities with their positions in the SMC, in the context of the galaxy formation, and
its chemical and dynamical evolution. Our goal is to investigate whether star clusters
projected toward tidally perturbed/induced SMC regions already formed therein, and from the
confirmed ones to assess whether these regions are detachments of the old outer SMC disk
or have been formed recently. 

\begin{figure*}
\includegraphics[width=\textwidth]{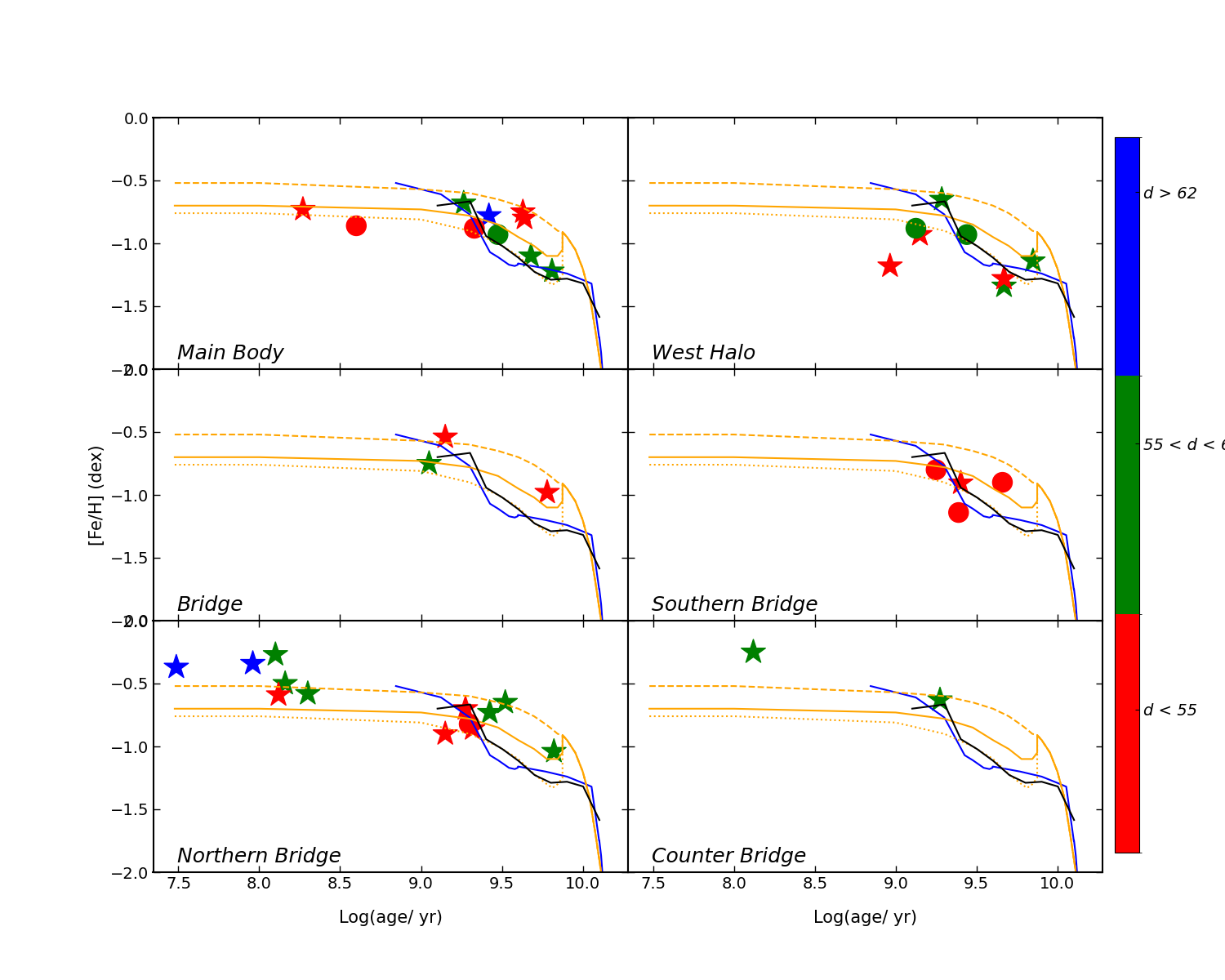}
\caption{Age-metallicity relationship for different SMC regions. 
Symbols are as in Figure~\ref{fig7}. The observed age-metallicity relationship
of \citet{pg13}, and the theoretical ones computed by \citet{pt1998} and
\citet{tb2009} are superimposed with black, blue, and orange lines,
respectively. The dashed and solid orange lines correspond to cases without
merger and with a merger of two similar mass galaxies, respectively.}
\label{fig8}
\end{figure*}

Figure~\ref{fig6} shows the variation of the derived metallicities with the cluster
heliocentric distances. For the sake of the reader, we represented the SMC Main Body boundaries 
with two vertical dashed lines. They come from the best-fitted rotation disk of SMC star clusters
obtained by \citet{piatti2021kinematics}. As can be seen,  there is a remarkable percentage of the
studied star cluster sample that is located beyond those limits, preferentially toward
closer distances. The derived distances uncover a spatial distribution of SMC star
clusters, in agreement with other independent studies that support the stripping scenario,
originated by the tidal interaction with the Large Magellanic Cloud (LMC) 
\citep{subramanianetal2017,cullinane2023magellanic,nidever2024}.  SMC outer disk star clusters 
have long been commonly thought to be old and hence, to be key to reconstructing
the early galaxy formation and chemical enrichment history, while young
star clusters are expected to be found in the inner galaxy region, where still remain
gas and dust. From these points of view, the older and more metal-poor star clusters
located in the outer SMC disk, as well as those younger and more metal-rich placed
in the inner galaxy, should not call our attention. However, the two closest star clusters
to the Sun (H86-97 and B99) and two other clusters located behind the SMC Main Body (L95 and HW64)
are young star clusters, which strikes our understanding about their roles in the galaxy formation 
and evolution processes. Notably, both young star clusters located behind the SMC formed
from a more chemically enriched material than their counterparts closer to the Sun.
Nevertheless, for the bulk of the studied star clusters, the younger a star cluster 
the more metal-rich, in agreement with the observed SMC cluster age-metallicity relationship 
\citep{piatti2011}.

\citet{carpinteroetal2013} modeled the dynamical interaction between the SMC and the LMC 
and their corresponding stellar cluster populations. Their simulations probing a wide range of 
parameters for the orbits of both galaxies showed that approximately 15 per cent of the SMC 
clusters are captured by the LMC. In addition, another 20-50 per cent of its clusters are ejected 
into the intergalactic medium. These results pave our understanding of the observed spatial 
distribution of the studied star clusters as objects that could have been reached by the
tidal interactions between both Magellanic Clouds. Indeed, The LMC is located closer to the Sun 
than the SMC, so that some SMC star clusters with heliocentric distances smaller than the
mean SMC distance could have experienced stripped from the LMC. Accompanying evidence of such an 
effect was found by \citet{pl2022}, who analyzed SMASH \citep{nideveretal2021} and {\it Gaia} EDR3
\citep{gaiaetal2016,gaiaetal2020b} data of the recently discovered star cluster YMCA-1, to 
conclude that it formed in the SMC
and then stripped by the LMC. YMCA-1 is a 9.6 Gyr old and moderately metal-poor ([Fe/H] = -1.16 dex)
star cluster located at 60.9 kpc from the Sun, and at $\sim$ 17.1 kpc to the East from the LMC center.

\citet{dias2014self} and \citet{dias2016smc} defined different zones throughout the outermost 
SMC regions, called: Northern Bridge, Counter Bridge, West Halo, Southern Bridge and Wing/Bridge, 
respectively \citep[see Figure~1 in][]{dias2022viscacha}. For comparison purposes, we have 
superimposed them in Figure~\ref{fig7}. As can be seen, all of them are populated by
star clusters analyzed in this work. In the Figure we have distinguished star clusters with
or without previous heliocentric distance estimates using filled stars and circles, respectively.
We employed the heliocentric distances of the star clusters as indicators of the 
tidally perturbed/induced origin of the outermost SMC regions where they appear, and
considered for our analysis three different distance cuts, namely. star clusters located closer than
55 kpc, between 55 kpc and 62 kpc, and beyond 62 kpc, according to \citet{piatti2021kinematics}
(see also Figure~\ref{fig6}). 

The Northern and Southern Bridges and the 
Wing/Bridge regions contain studied star clusters located in front of the SMC Main Body, 
which support their tidal origin. These regions are facing the LMC, which is at $\sim$ 50 kpc
from the Sun \citep{dgetal14}, so that it would seem that exists a vast region between both galaxies 
connecting them. This outcome is in very good agreement with the results obtained by
\citet{wagner2017properties}, who derived metallicity and distance
distributions of RR Lyrae variables stars to probe the structure of the Magellanic system 
as a whole, revealing a smooth transition that connects the galaxies.
Star clusters closer than 55 kpc from the Sun are also seen toward the
SMC Main Body, inside an ellipse of 2$\degr$. These star clusters confirm the larger extension 
of  the galaxy along the line-of-sight, with a 1:2:4 3D shape, being the declination, right 
ascension and line-of-sight the three axes, respectively \citep{ripepi2017vmc,muravevaetal2018}. 
Likewise, star clusters farther than 62 kpc from the Sun are observed in the Northern Bridge 
and projected toward the SMC Main Body.

Besides the cluster heliocentric distances, the cluster ages and metallicities also play a role
that sheds light on to the origin of the SMC stellar structures where the clusters belong to.
In this sense, comparing the SMC age-metallicity relationship (AMR) with those of the 
aforementioned outermost regions contribute to understand at what extend these regions have 
chemically evolved differently from the SMC. For comparison purposes, we used the
AMR derived by \citet{pg13} from Washington photometry of field stars distributed throughout
the whole galaxy, and the theoretical ones computed by \citet{pt1998} and 
\citet{tb2009}, respectively. \citet{pt1998} predicted an intensive star formation and
chemical enrichment during the SMC formation epoch and a rapid burst of chemical
enrichment about 3 Gyr ago. On the other hand, \citet{tb2009} predicted a major merger at 
$\sim$ 7.5 Gyr ago between two small galaxies of similar mass.
Figure~\ref{fig8} depicts the resulting AMR for different outermost SMC regions with 
star clusters colored according to their heliocentric distances. 

In general, the studied clusters located in different SMC regions follow the overall
SMC AMR. The most discordant star cluster would seem to be L13, located in the West Halo
(log(age /yr) = 8.97, [Fe/H] = -1.18 dex). Its relatively low metal content could suggest
that it formed from not-well mixed gas. Star clusters located at heliocentric distances $<$
55 kpc do not distinguish chemically from those populating the SMC Main Body. From this 
outcome we speculate with the possibility that they could have formed in the SMC and then 
migrated beyond the galaxy periphery dragged by the tidal effects caused by the 
LMC/SMC interaction. In addition, relatively young and distant clusters
(55 kpc $>$ $d$ $>$ 62 kpc) located in the Northern Bridge tell us that enriched gas has also
reached the outermost SMC regions.

\section{Conclusions}

The SMC is known to have experienced tidal effects from its interaction with the LMC.
As expected, its star cluster population has not been indifferent to such an interaction 
history, which can be deciphered from the relationship between their ages, metallicities 
and distances. To this respect, we selected 40 SMC star clusters aiming at deriving their
astrophysical properties, from which to look for the connection between the star clusters
and the tidally perturbed/induced outermost SMC regions where they are located. 
We used relatively deep cluster CMDs built from SMASH DR2 photometry and, once they were
properly cleaned from the field star contamination, we derived their cluster fundamental
parameters. 

We found that the homogeneously derived clusters' ages, heliocentric distances and 
metallicities are put on a uniform scale validated by published reliable estimates.
We obtained for the first time heliocentric distances for 15 star clusters, which
represents an increase of $\sim$50 per cent of the number of SMC clusters pertaining to 
the group of clusters with derived distances \citep{piatti2023depth}; the SMC containing
more than 650 star clusters \citep{bicaetal2020}. This small sample of clusters with 
heliocentric distances are distributed throughout the SMC Main Body, similarly as field 
stars do. For this reason, finding star clusters beyond the SMC body triggers the speculation 
that they could have been stripped from the interaction with the LMC.

When dealing with the estimated cluster heliocentric distances, we found that nearly half 
of the studied star clusters are located beyond the SMC Main Body. Most of them are in front 
of the SMC, but there are some ones also located behind it. This result is not
unexpected, bearing in mind independent observational evidence of the tidal interaction
between the LMC and the SMC. Here we confirm the existence of that physical connection between 
both galaxies, using star clusters are tracers.

From the analysis of the AMRs of star clusters located in different outermost SMC regions, 
and in the Main Body of the galaxy, we found that intermediate-age and old star clusters
follow in general the overall galaxy chemical enrichment history. Since some of them
are placed at the present time beyond the SMC Main Body boundaries, we conclude that they
could have traveled outward under the effects of the tidal interaction between the Magellanic 
Clouds. Moreover, not only star clusters formed in the SMC have surpassed its outskirts,
but also chemically enriched gas. This is supported by the formation of young metal-rich
star clusters in the Northern Bridge that also span a wide range of heliocentric distances.
Therefore, the SMC has become a dwarf galaxy scarred by LMC tidal effects with
distance-perturbed  and newly induced (formed) outermost stellar structures.
Precisely, not considering such a complex galaxy reality could mislead our understanding 
of the formation and evolution of the SMC, for example, the present spatial distribution 
of the star cluster metal content.

\begin{acknowledgements}
We thank the referee for the thorough reading of the manuscript 
and timely suggestions to improve it.
This research uses services or data provided by the Astro Data Lab at NSF's National
 Optical-Infrared Astronomy Research Laboratory. NSF's OIR Lab is operated by the Association
 of Universities for  Research in Astronomy (AURA), Inc. under a cooperative agreement with
 the National Science 
Foundation.

Data for reproducing the figures and analysis in this work will be available upon request to
 the first author.

\end{acknowledgements}


\appendix
\section{Cleaned CMDs with the corresponding theoretical isochrones}

\begin{figure}[H]
\includegraphics[width=\columnwidth]{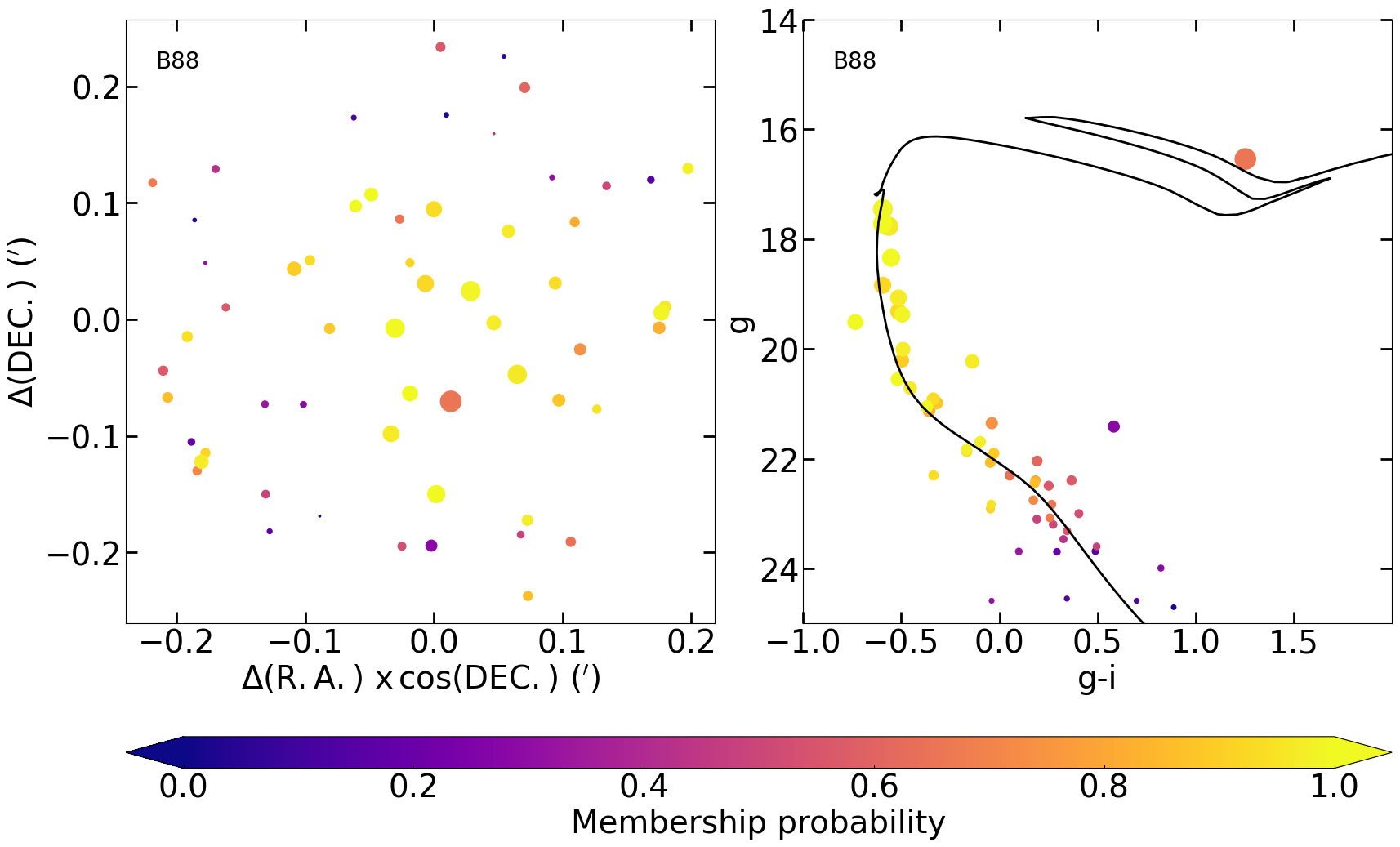}
\end{figure}
\begin{figure}[H]
\includegraphics[width=\columnwidth]{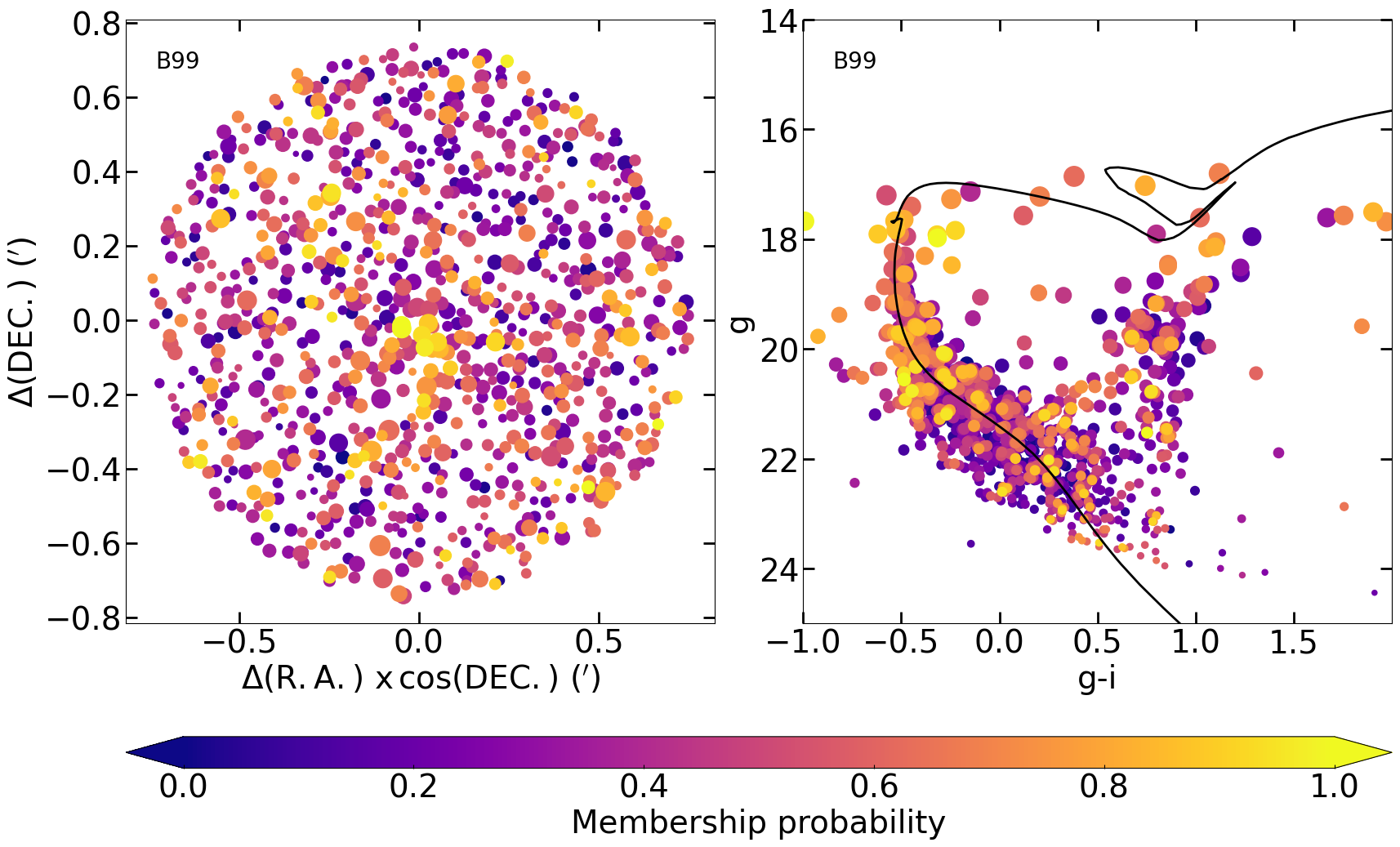}
\end{figure}
\begin{figure}[H]
\includegraphics[width=\columnwidth]{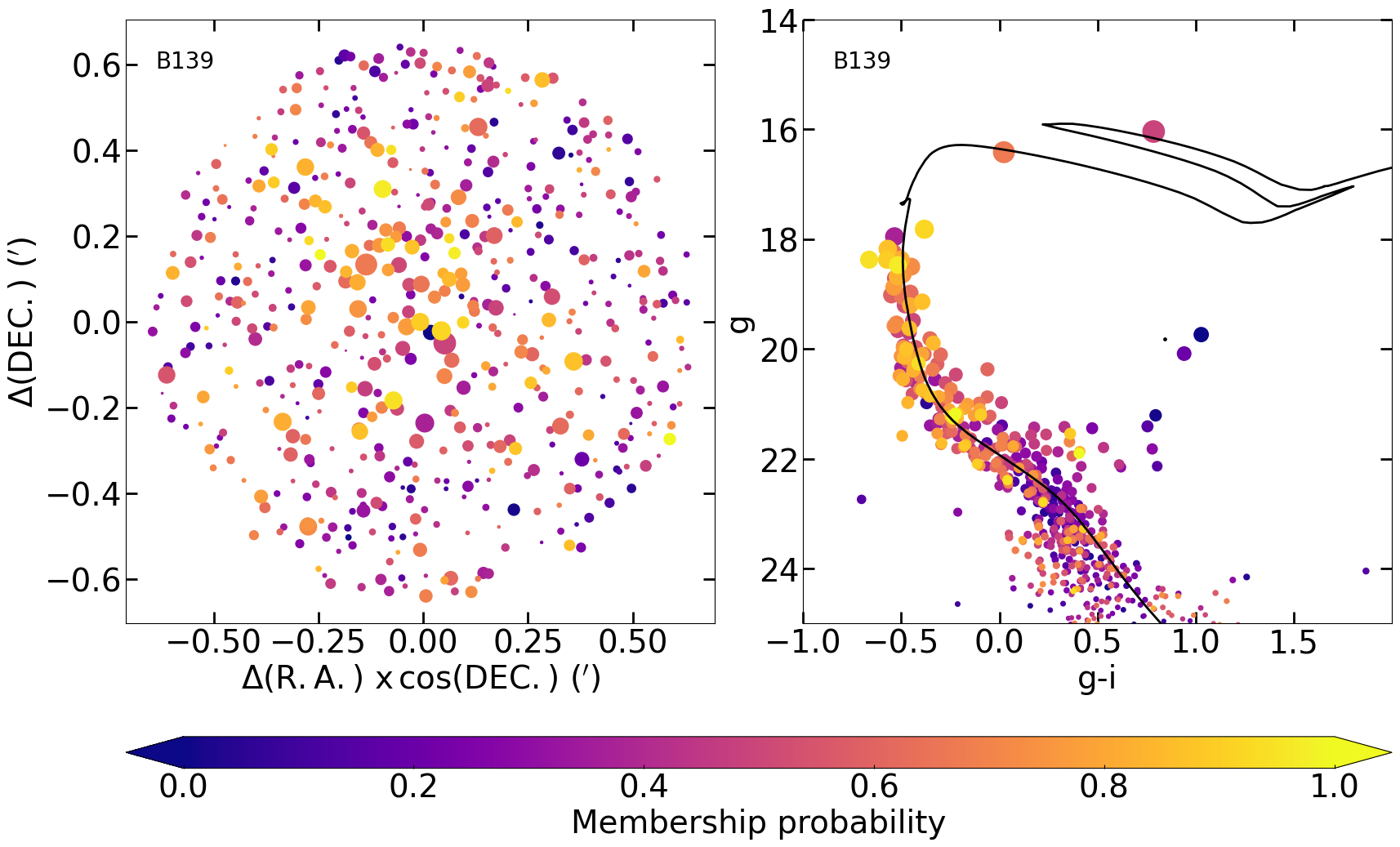}
\label{fig:B139}
\end{figure}
\begin{figure}[H]
\includegraphics[width=\columnwidth]{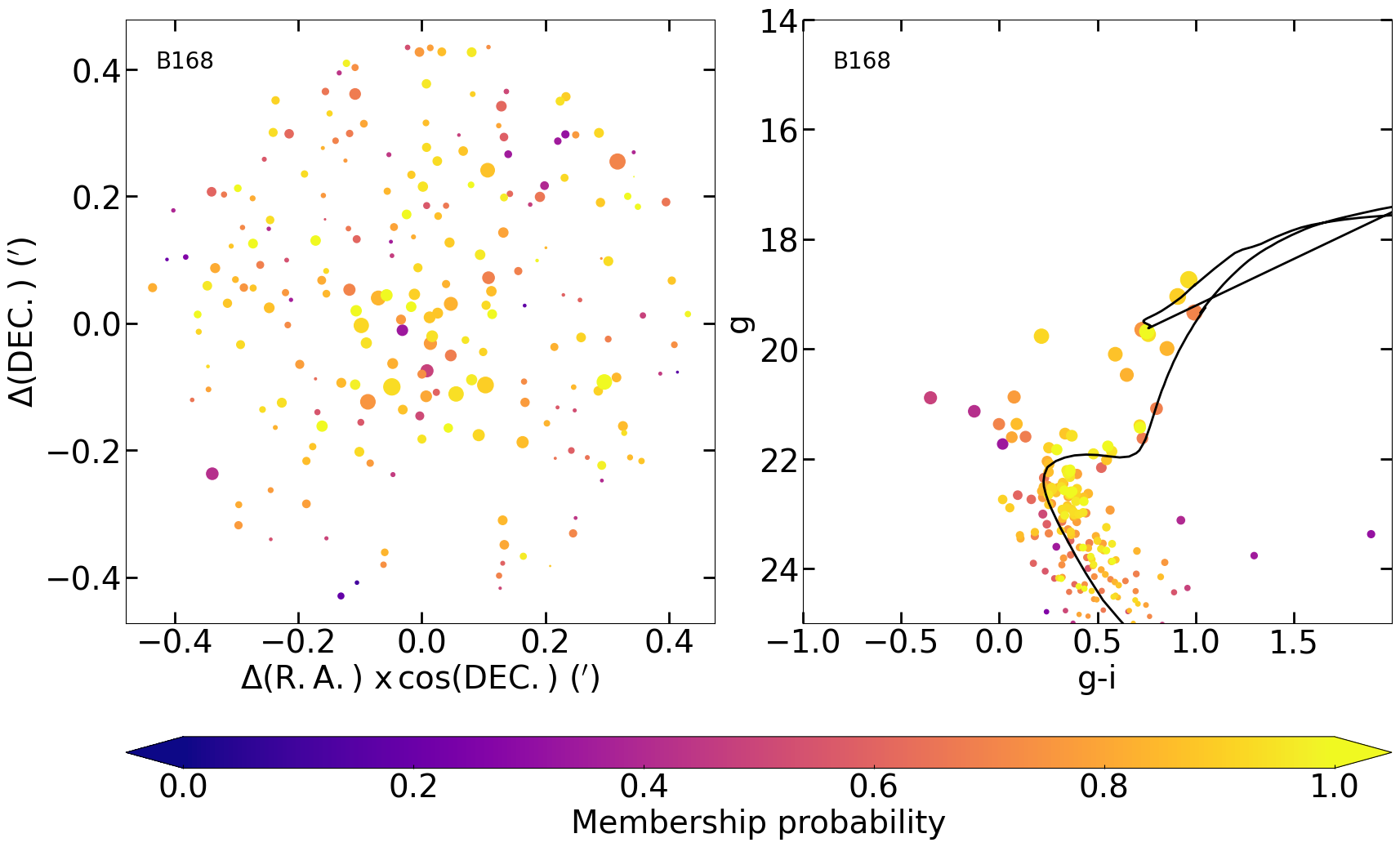}
\end{figure}
\begin{figure}[H]
\includegraphics[width=\columnwidth]{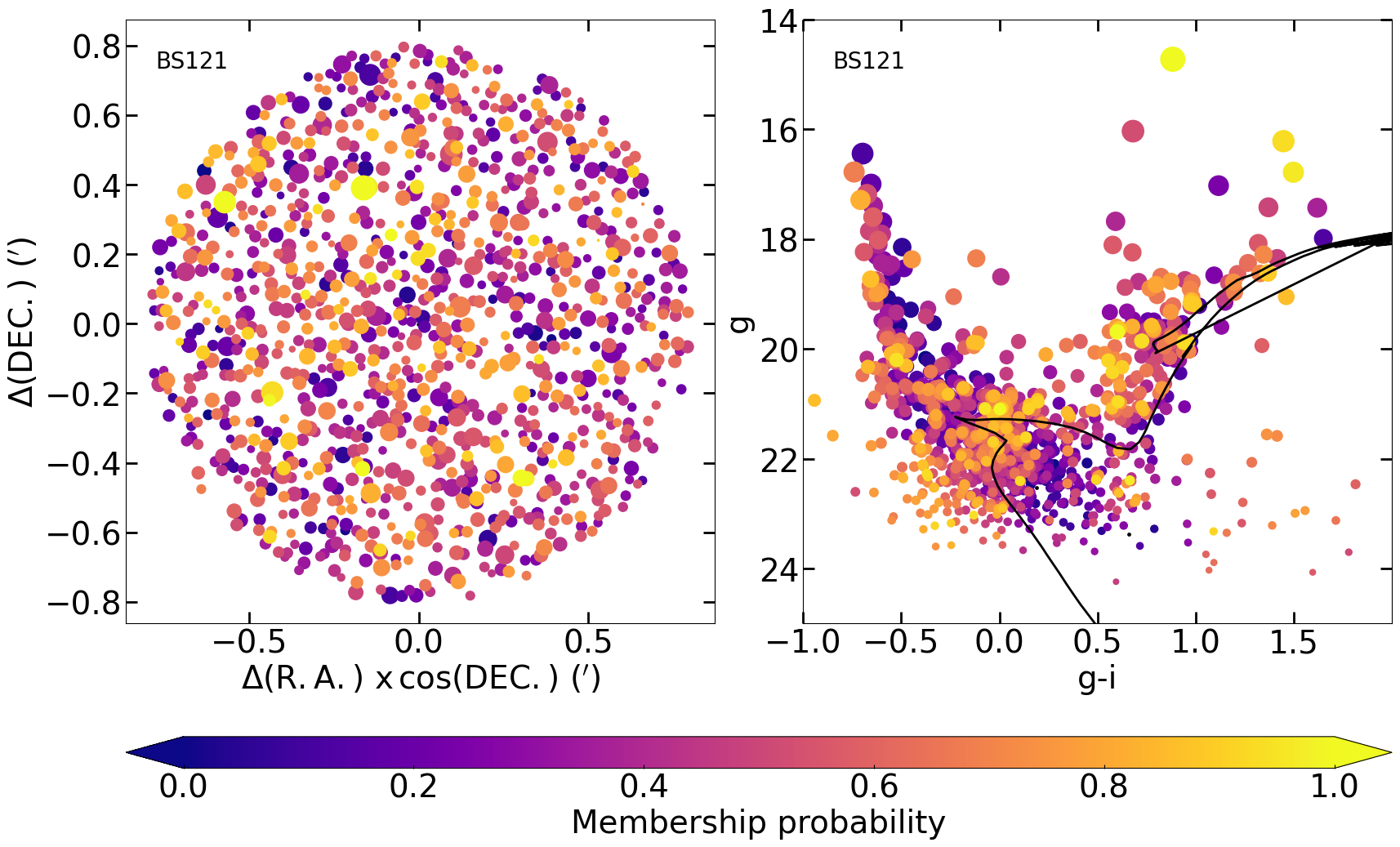}
\end{figure}
\begin{figure}[H]
\includegraphics[width=\columnwidth]{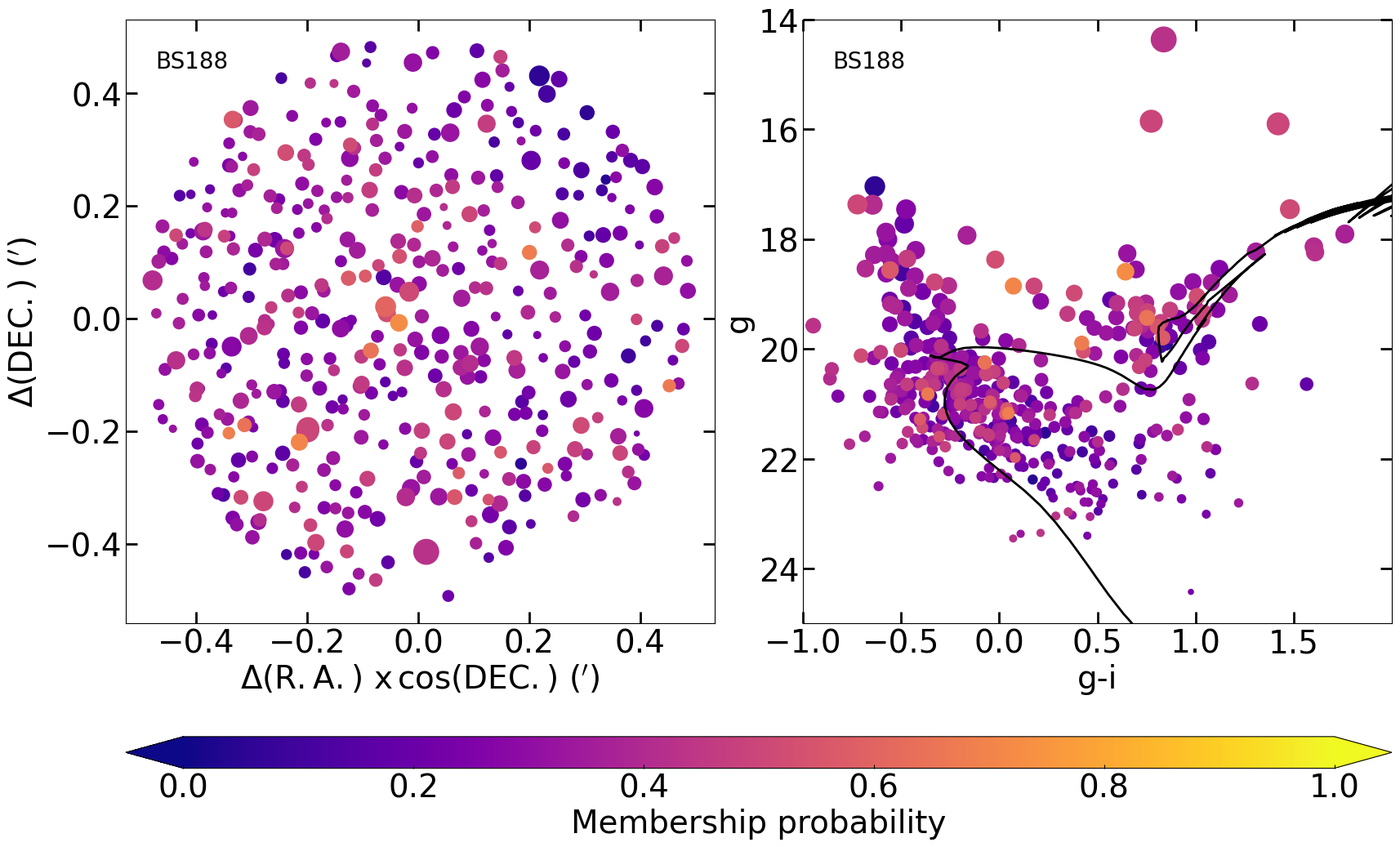}
\end{figure}
\begin{figure}[H]
\includegraphics[width=\columnwidth]{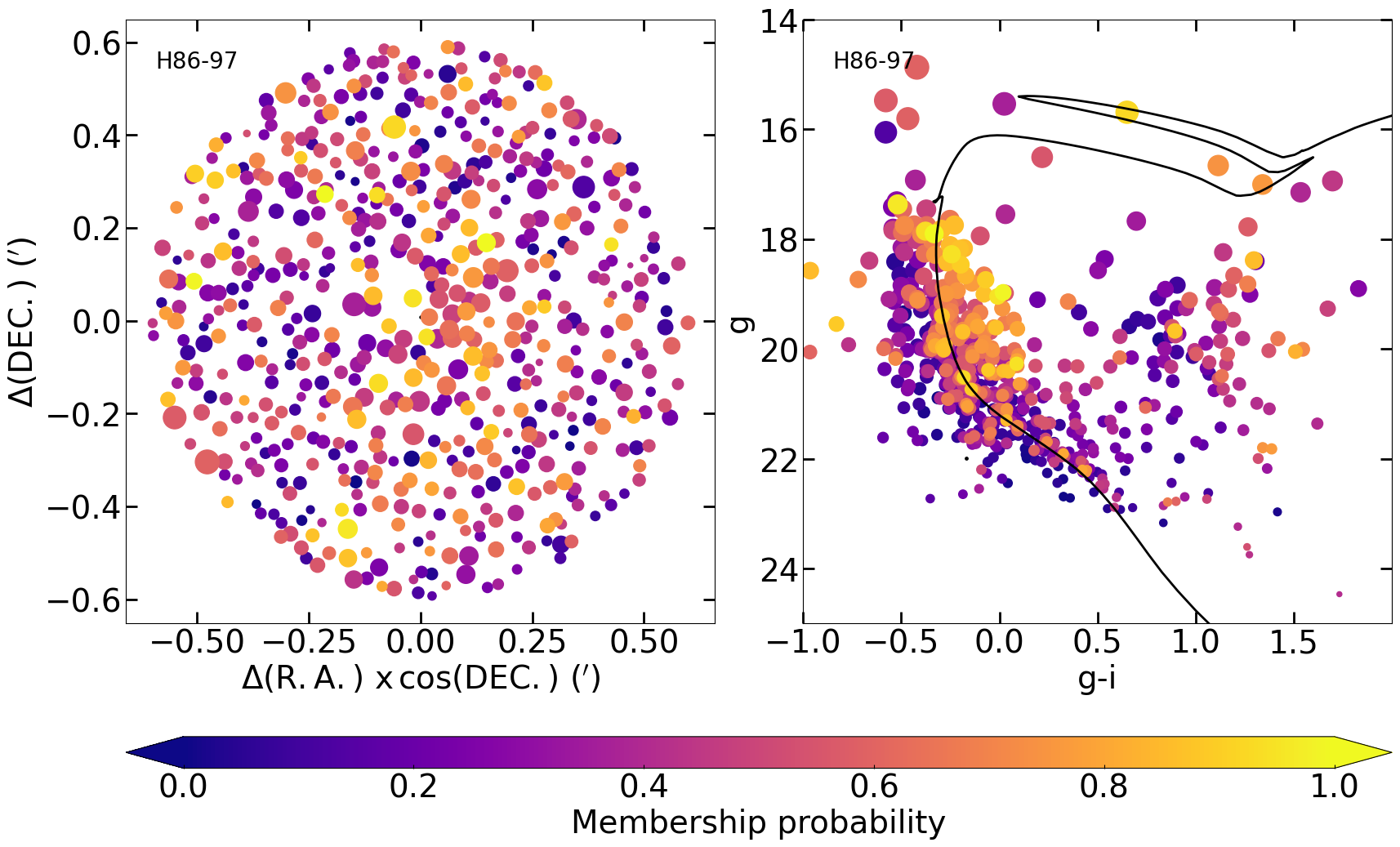}
\end{figure}
\begin{figure}[H]
\includegraphics[width=\columnwidth]{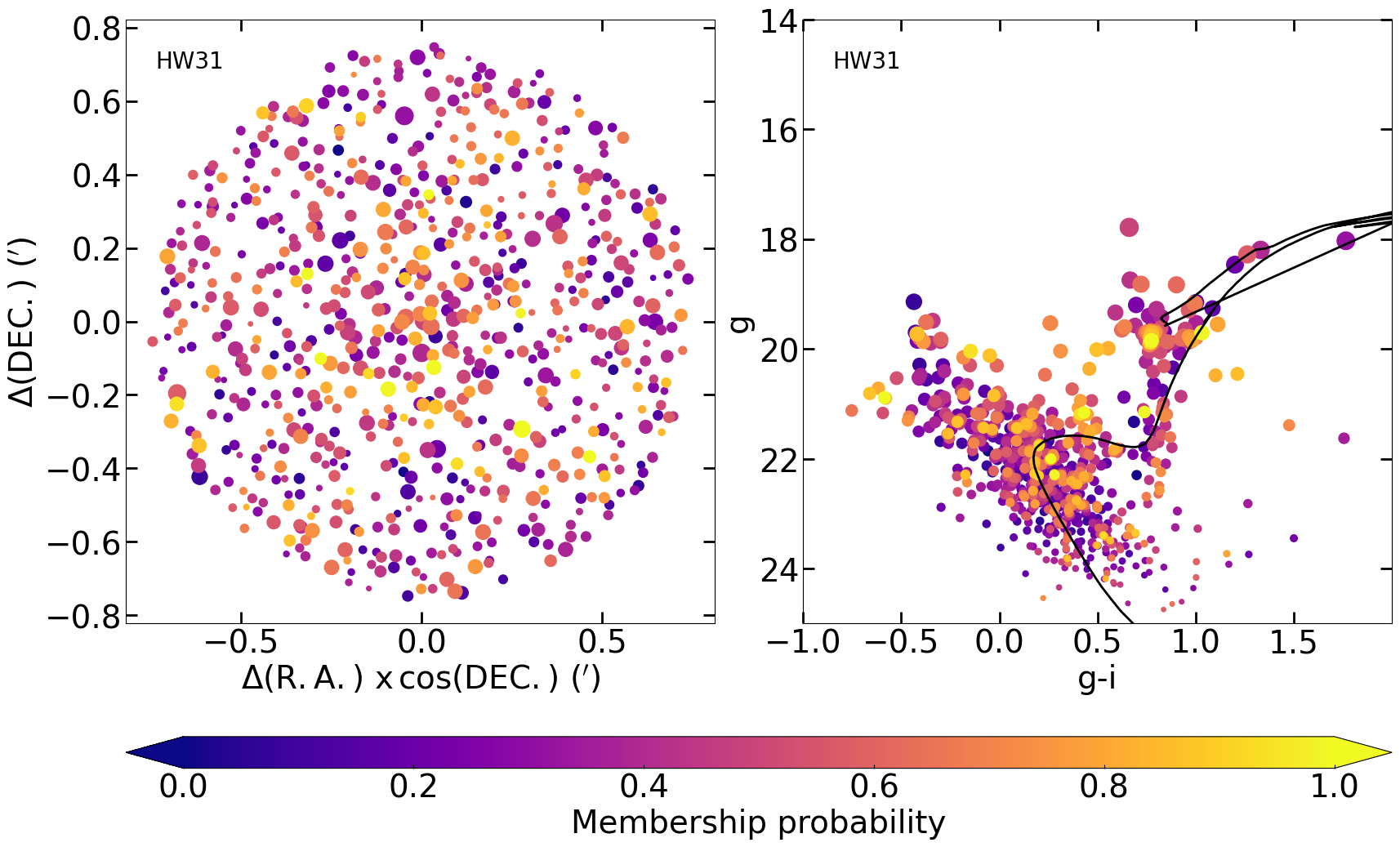}
\end{figure}
\begin{figure}[H]
\includegraphics[width=\columnwidth]{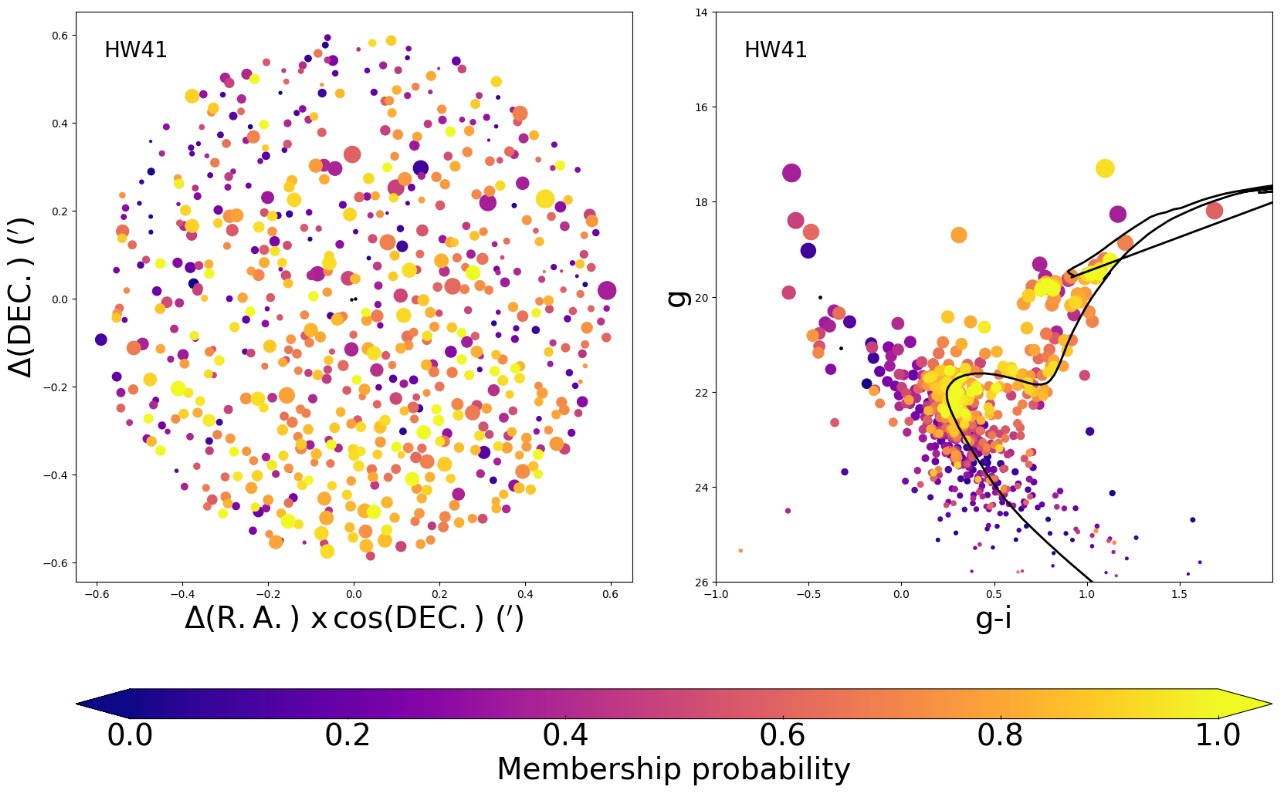}
\end{figure}
\begin{figure}[H]
\includegraphics[width=\columnwidth]{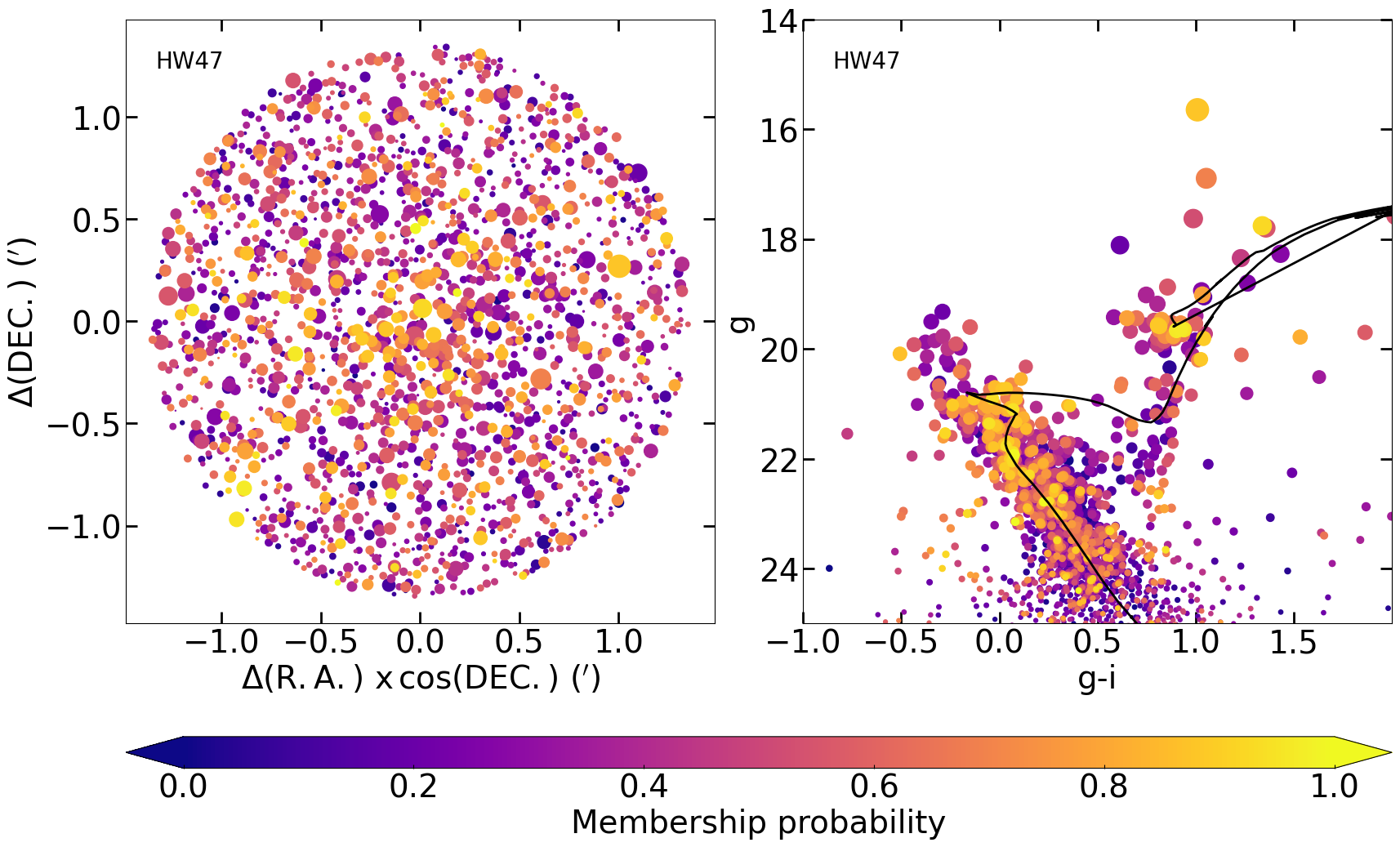}
\end{figure}
\begin{figure}[H]
\includegraphics[width=\columnwidth]{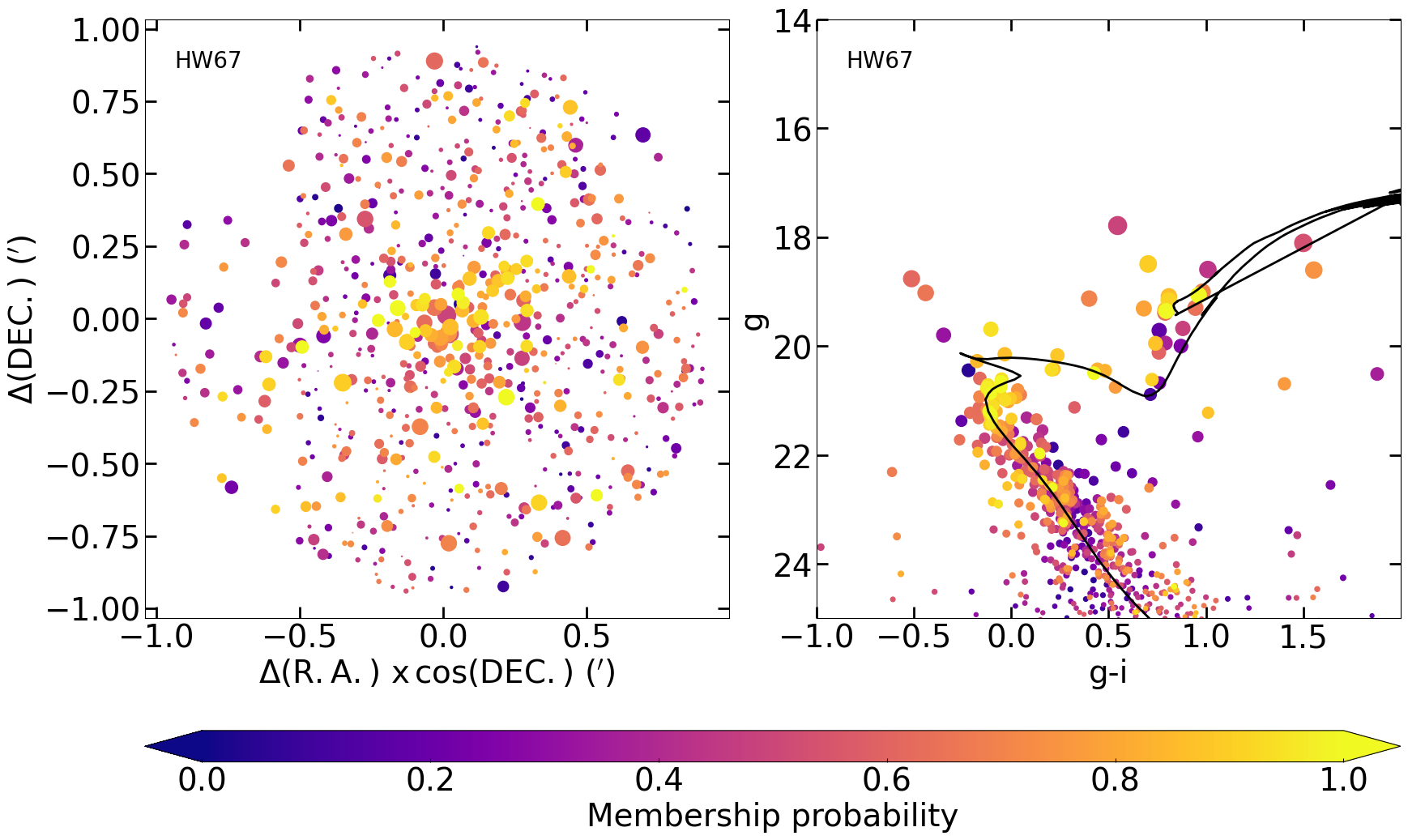}
\end{figure}
\begin{figure}[H]
\includegraphics[width=\columnwidth]{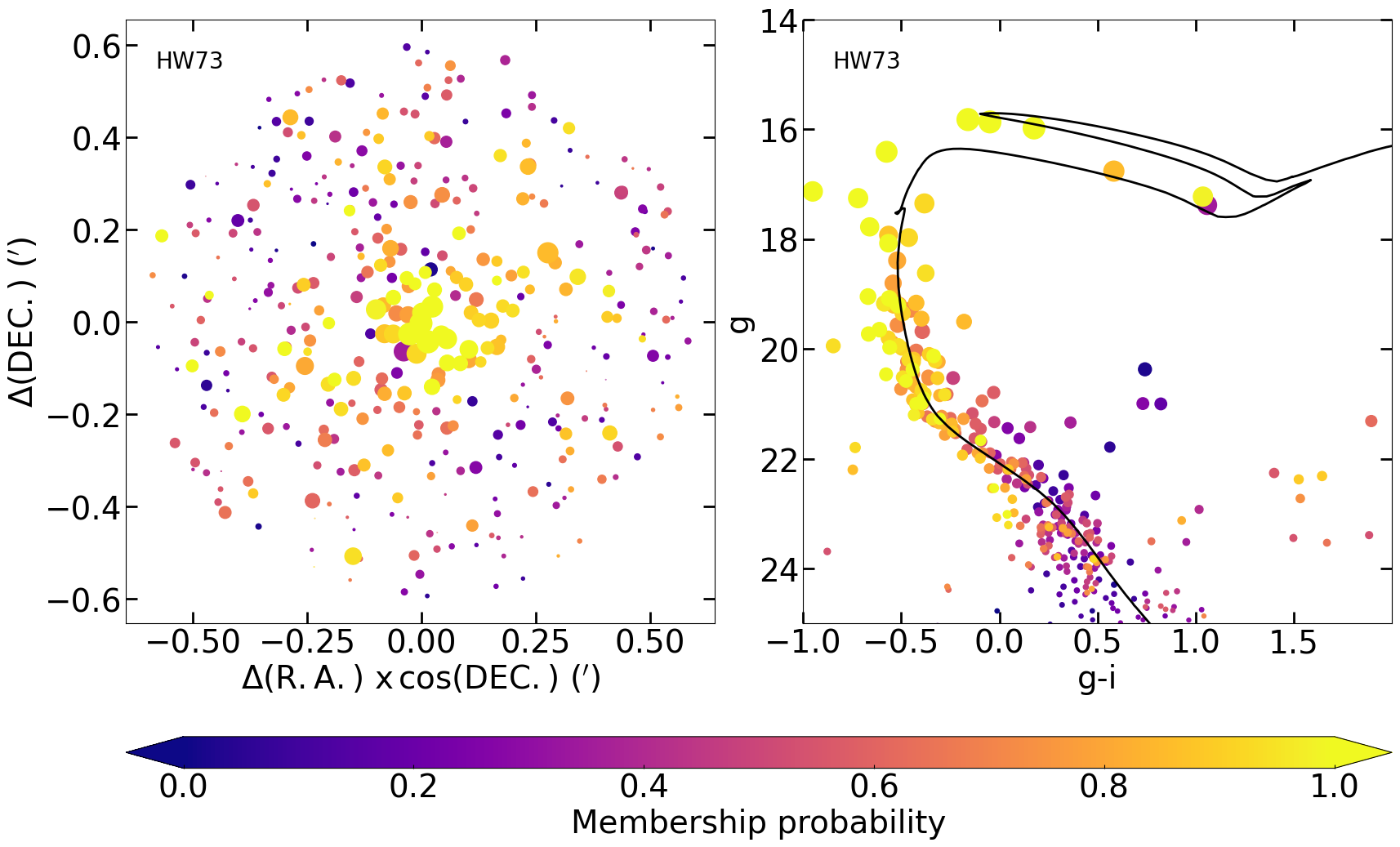}
\label{fig:HW73}
\end{figure}
\begin{figure}[H]
\includegraphics[width=\columnwidth]{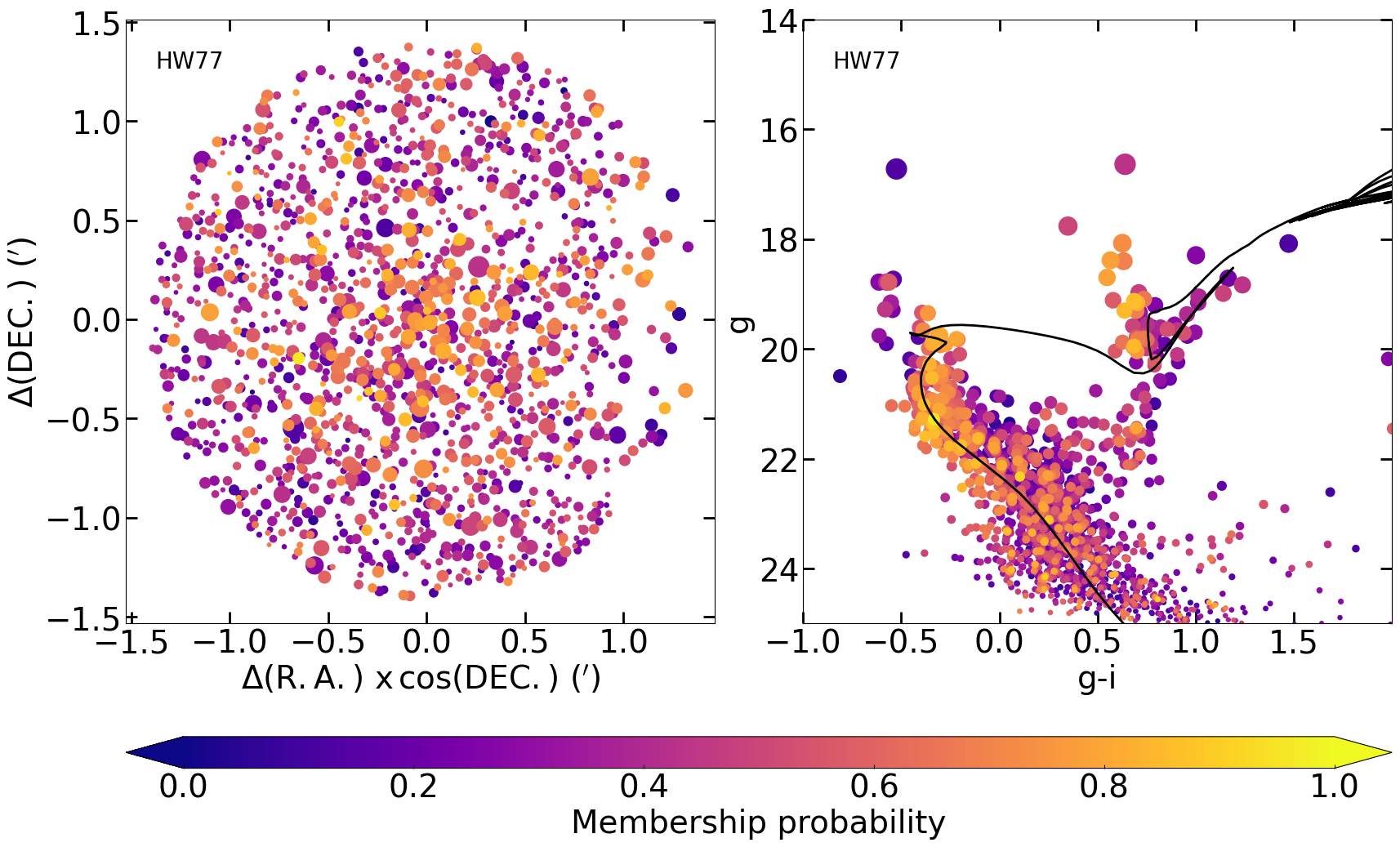}
\end{figure}
\begin{figure}[H]
\includegraphics[width=\columnwidth]{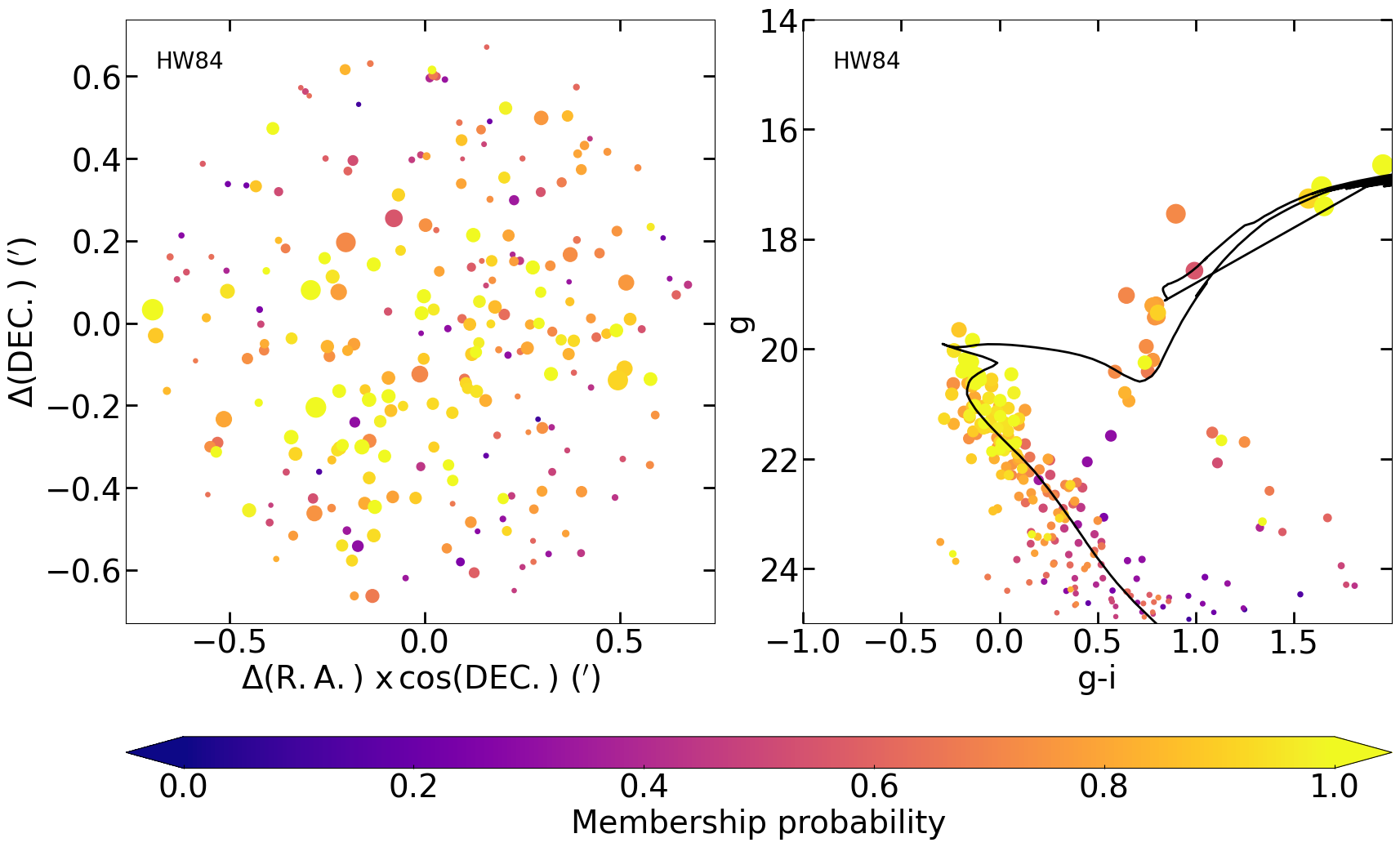}
\end{figure}
\begin{figure}[H]
\includegraphics[width=\columnwidth]{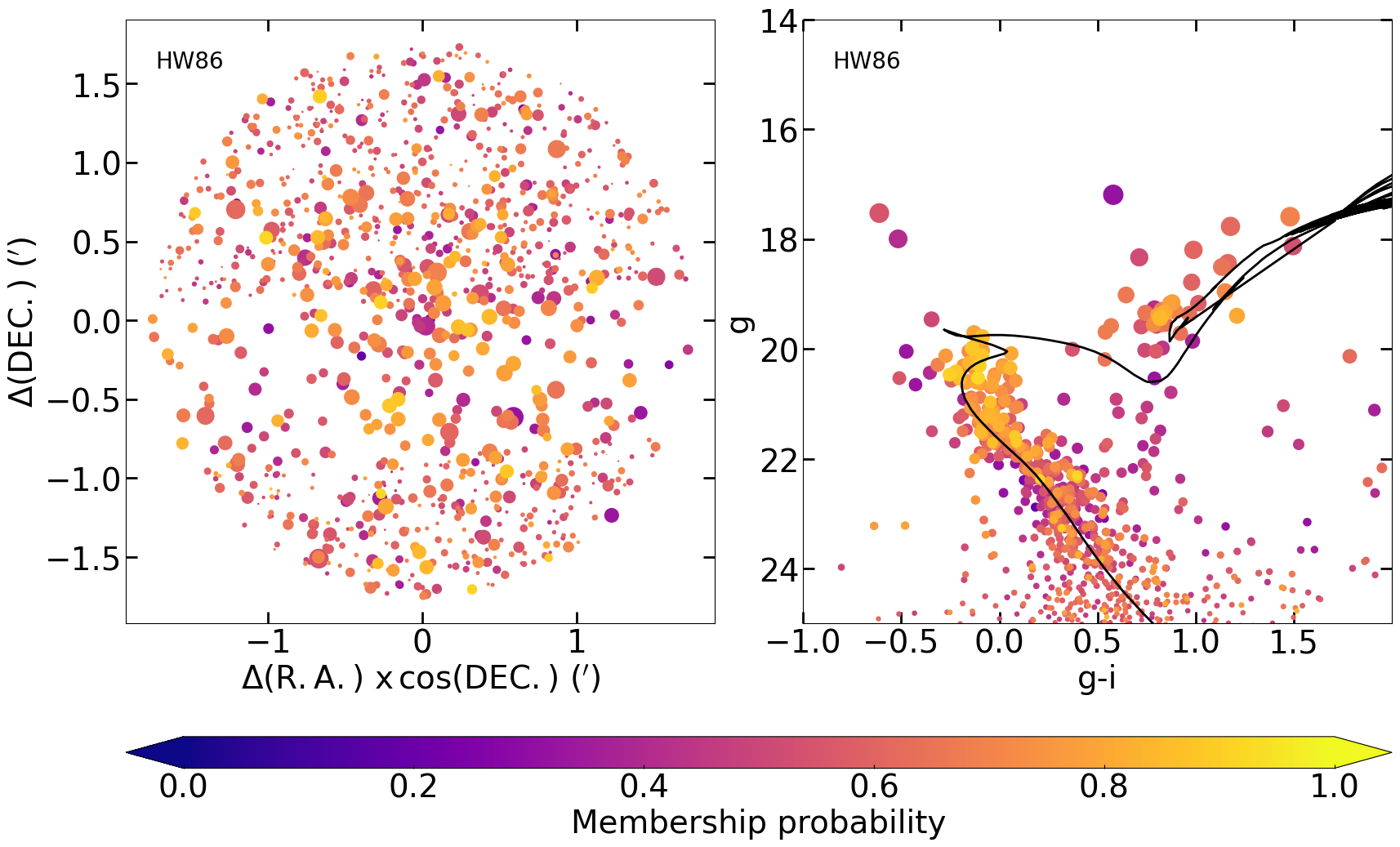}
\end{figure}
\begin{figure}[H]
\includegraphics[width=\columnwidth]{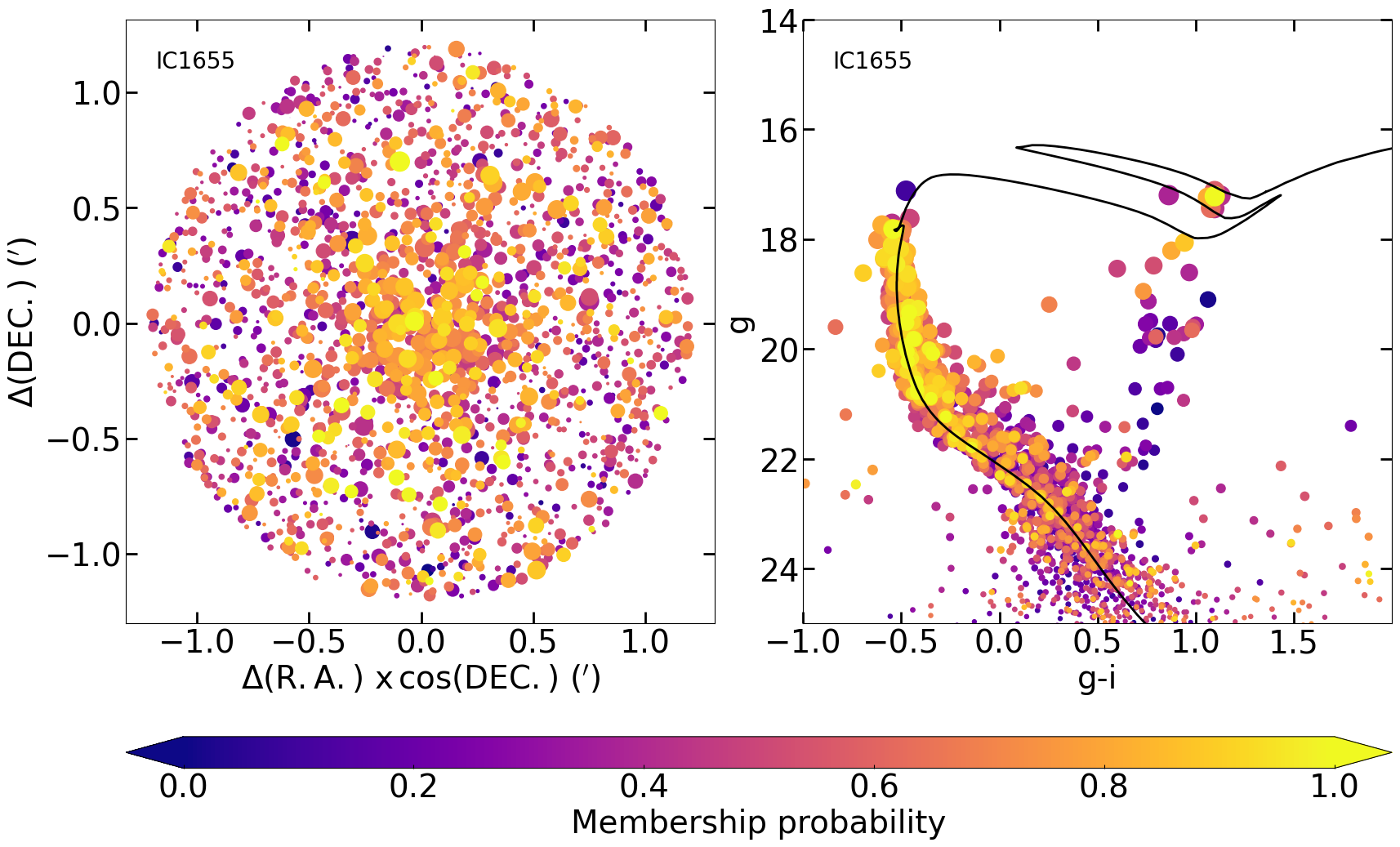}
\end{figure}
\begin{figure}[H]
\includegraphics[width=\columnwidth]{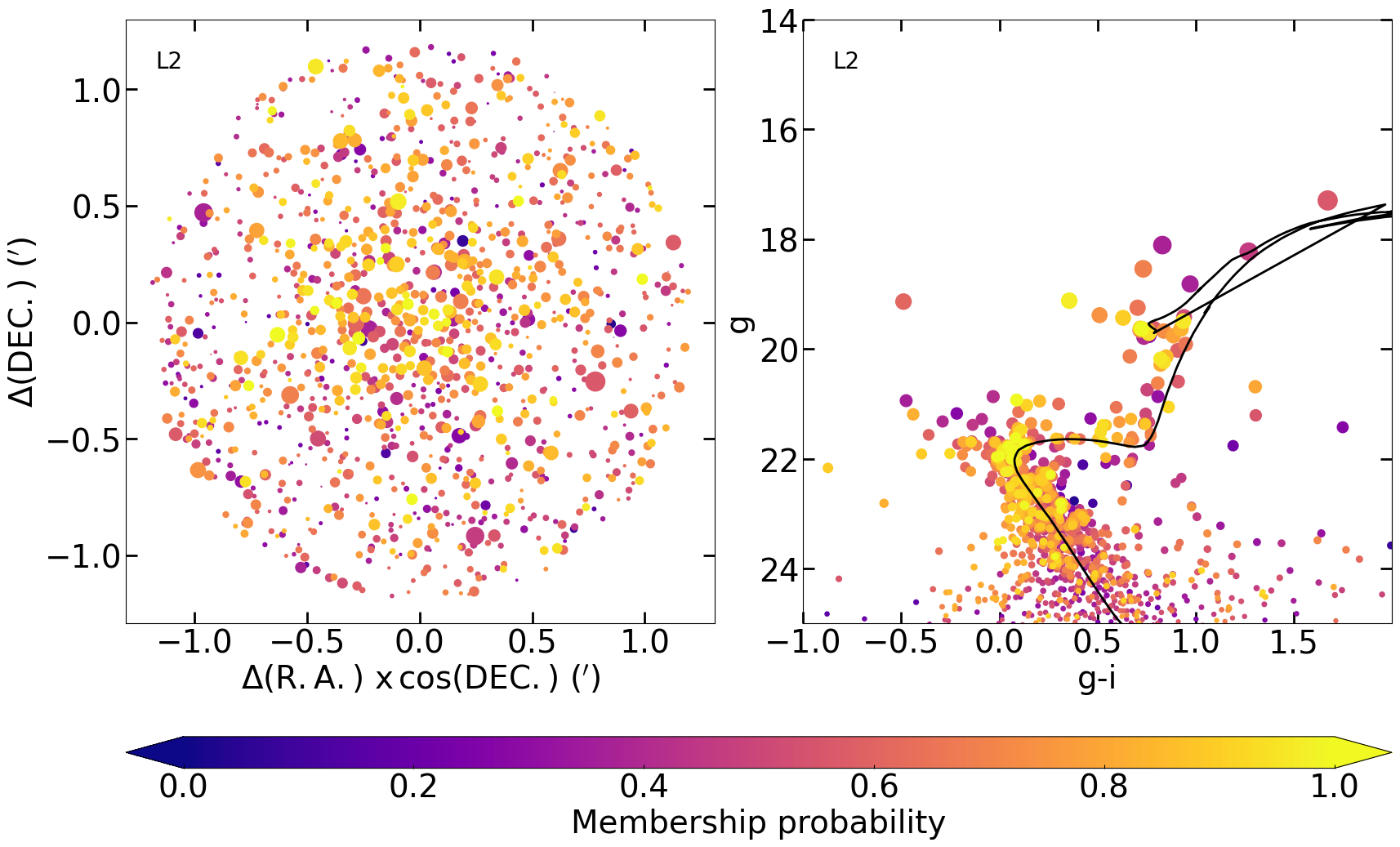}
\end{figure}
\begin{figure}[H]
\includegraphics[width=\columnwidth]{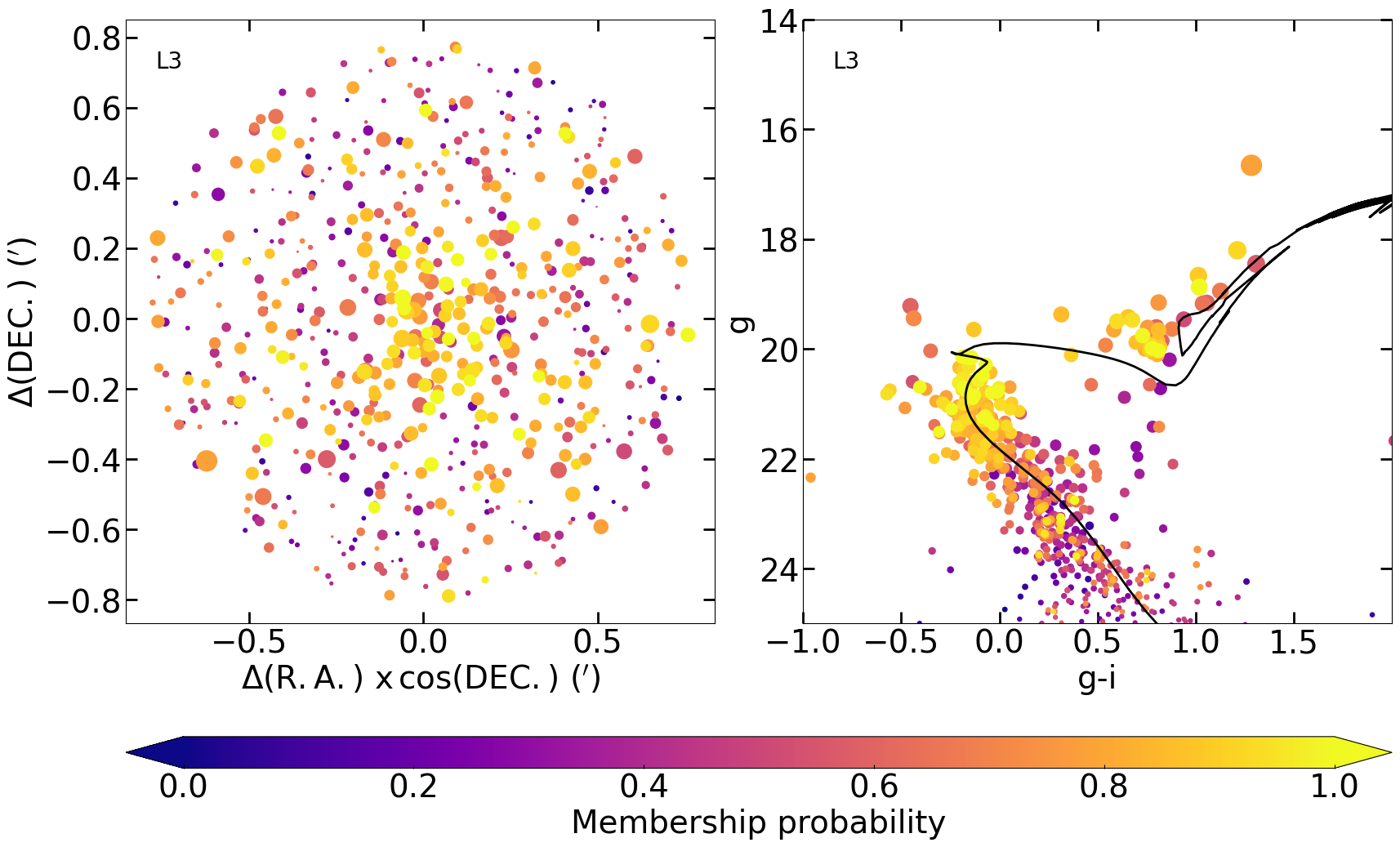}
\end{figure}
\begin{figure}[H]
\includegraphics[width=\columnwidth]{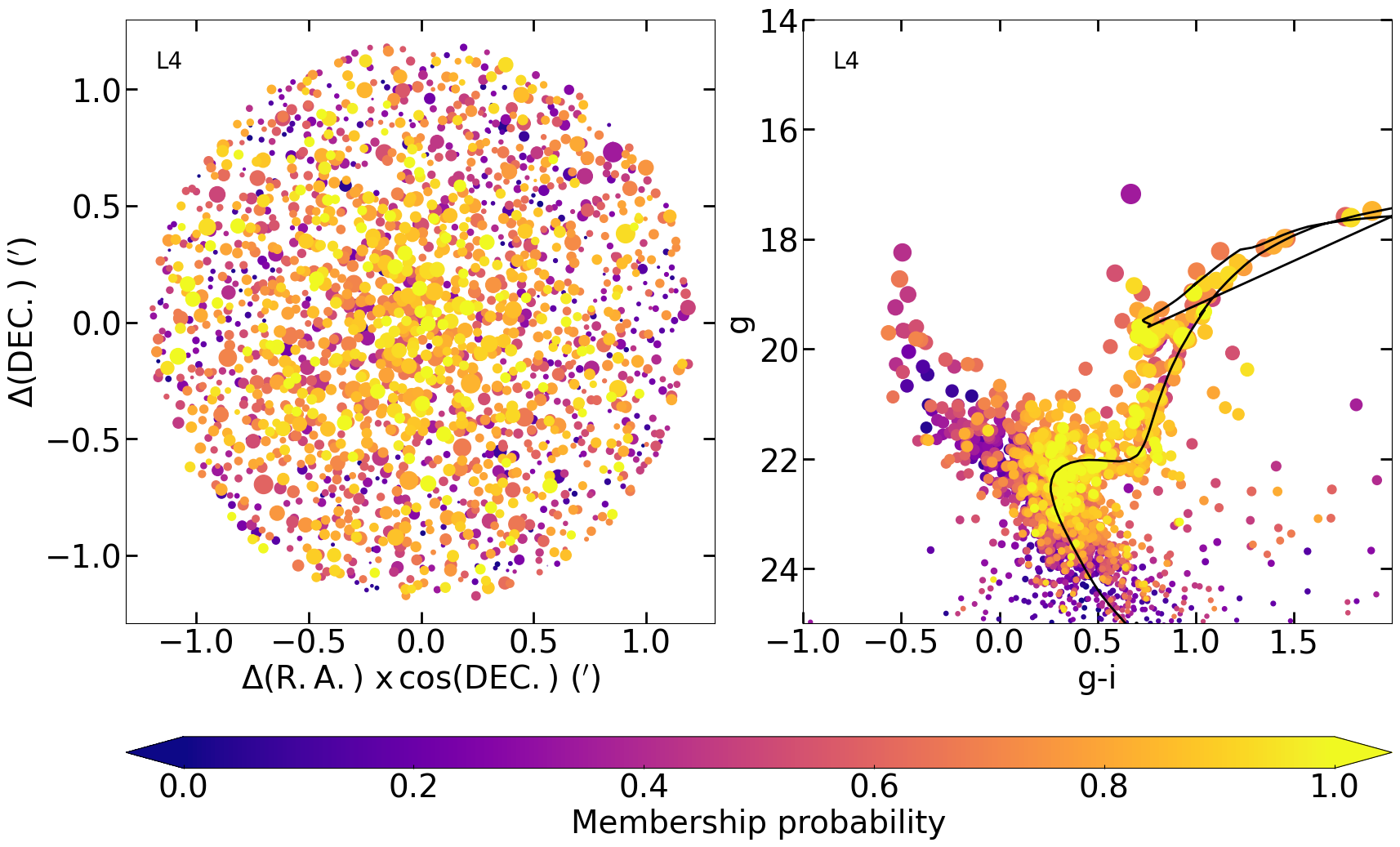}
\end{figure}
\begin{figure}[H]
\includegraphics[width=\columnwidth]{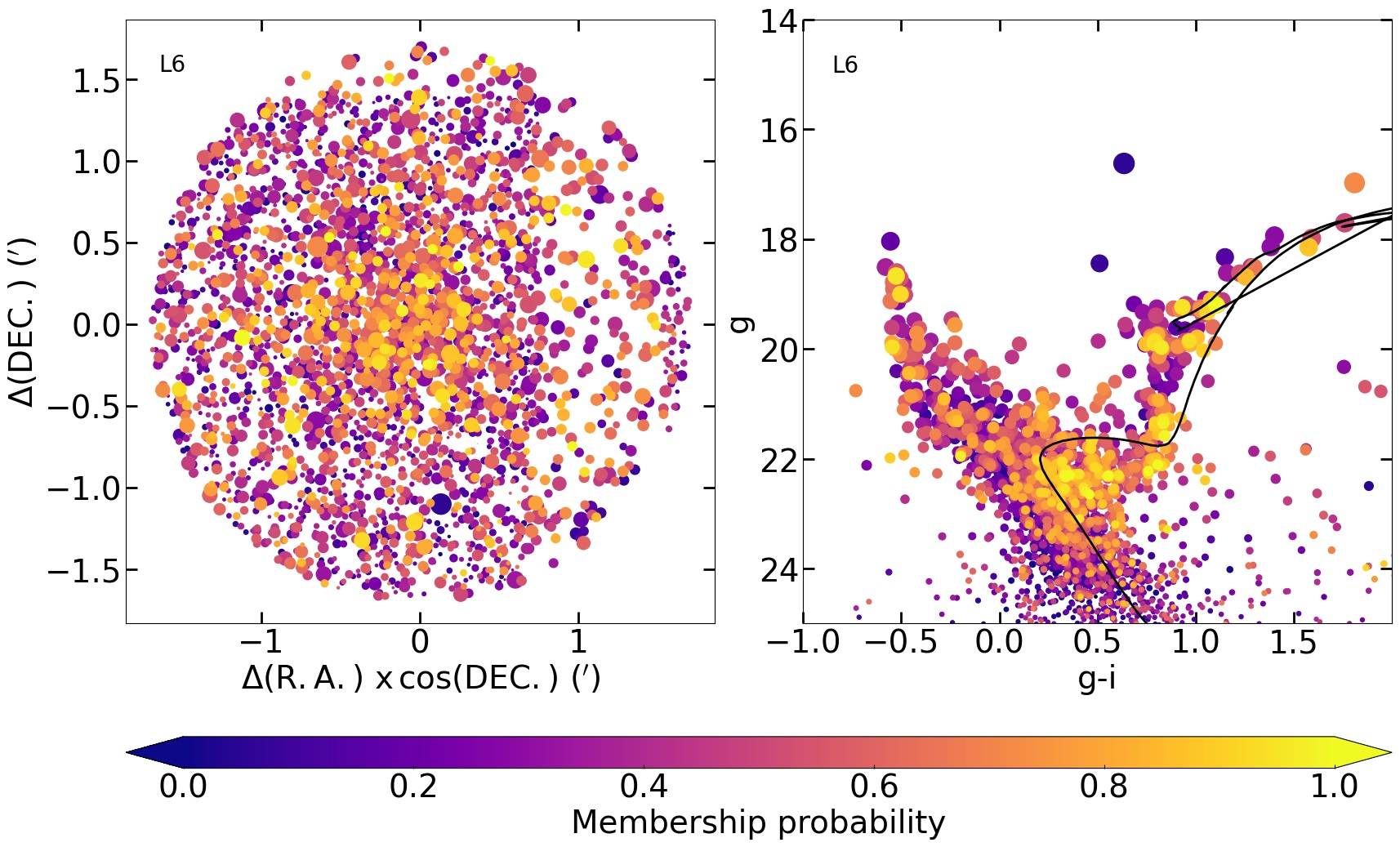}
\end{figure}
\begin{figure}[H]
\includegraphics[width=\columnwidth]{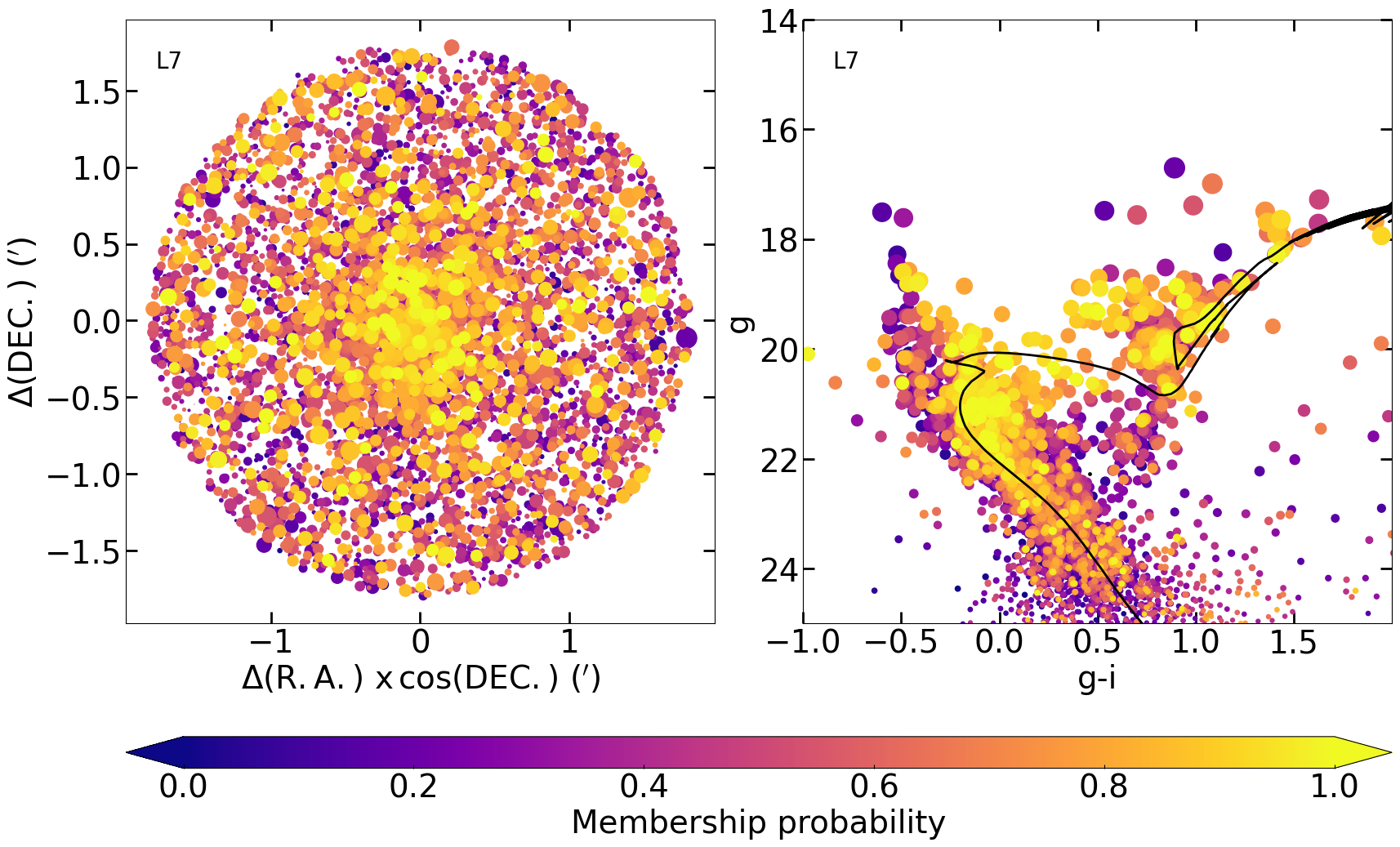}
\end{figure}
\begin{figure}[H]
\includegraphics[width=\columnwidth]{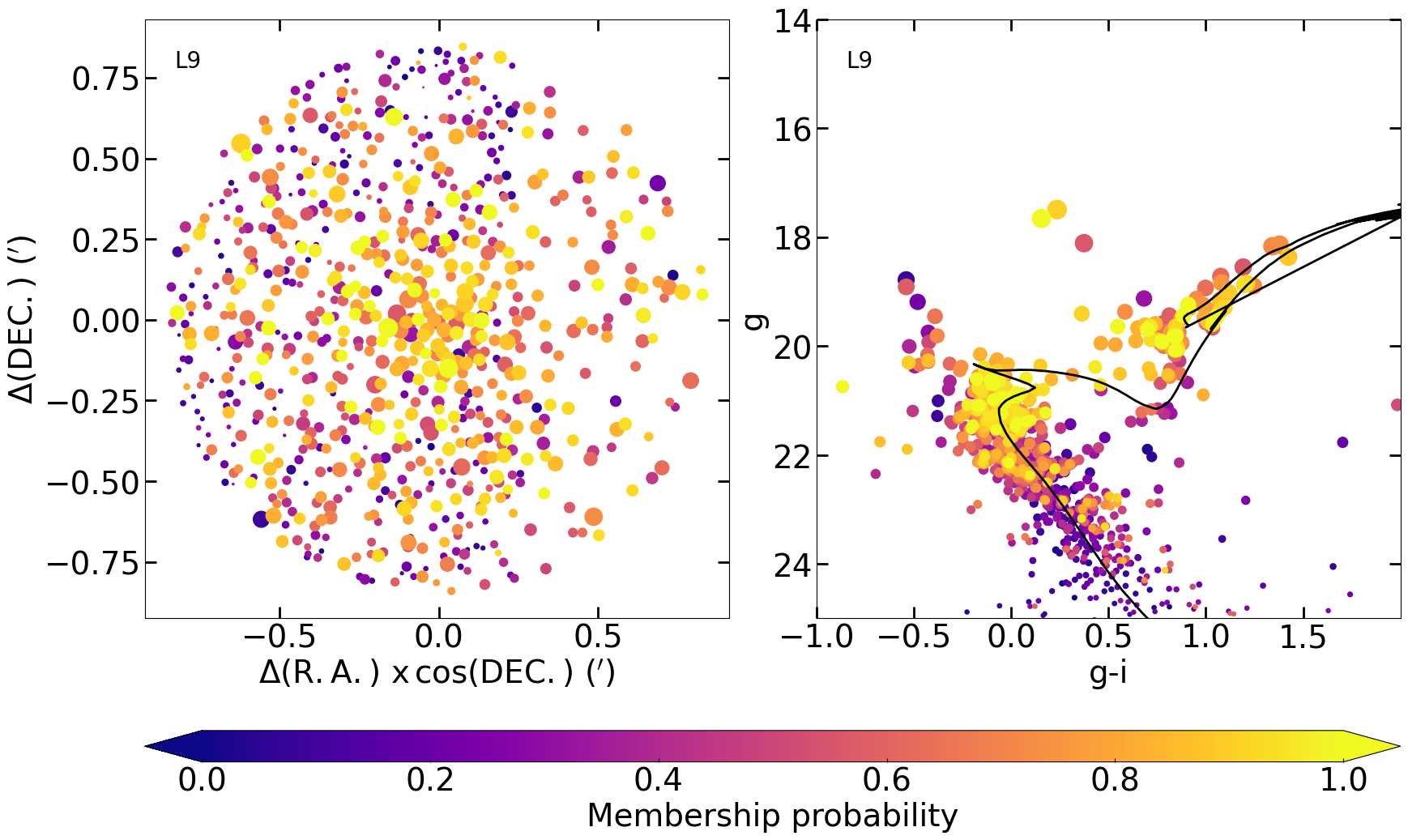}
\end{figure}
\begin{figure}[H]
\includegraphics[width=\columnwidth]{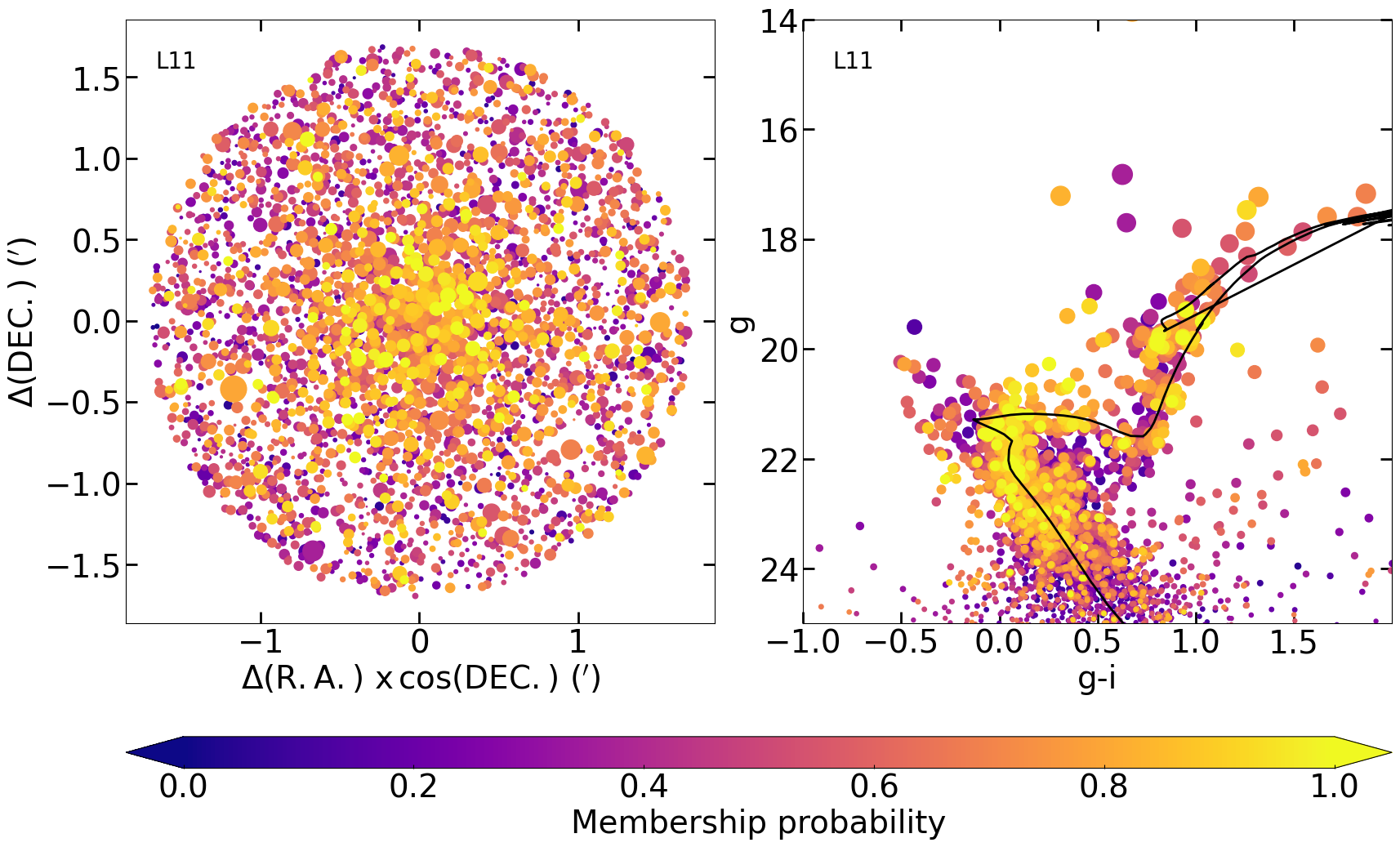}
\end{figure}
\begin{figure}[H]
\includegraphics[width=\columnwidth]{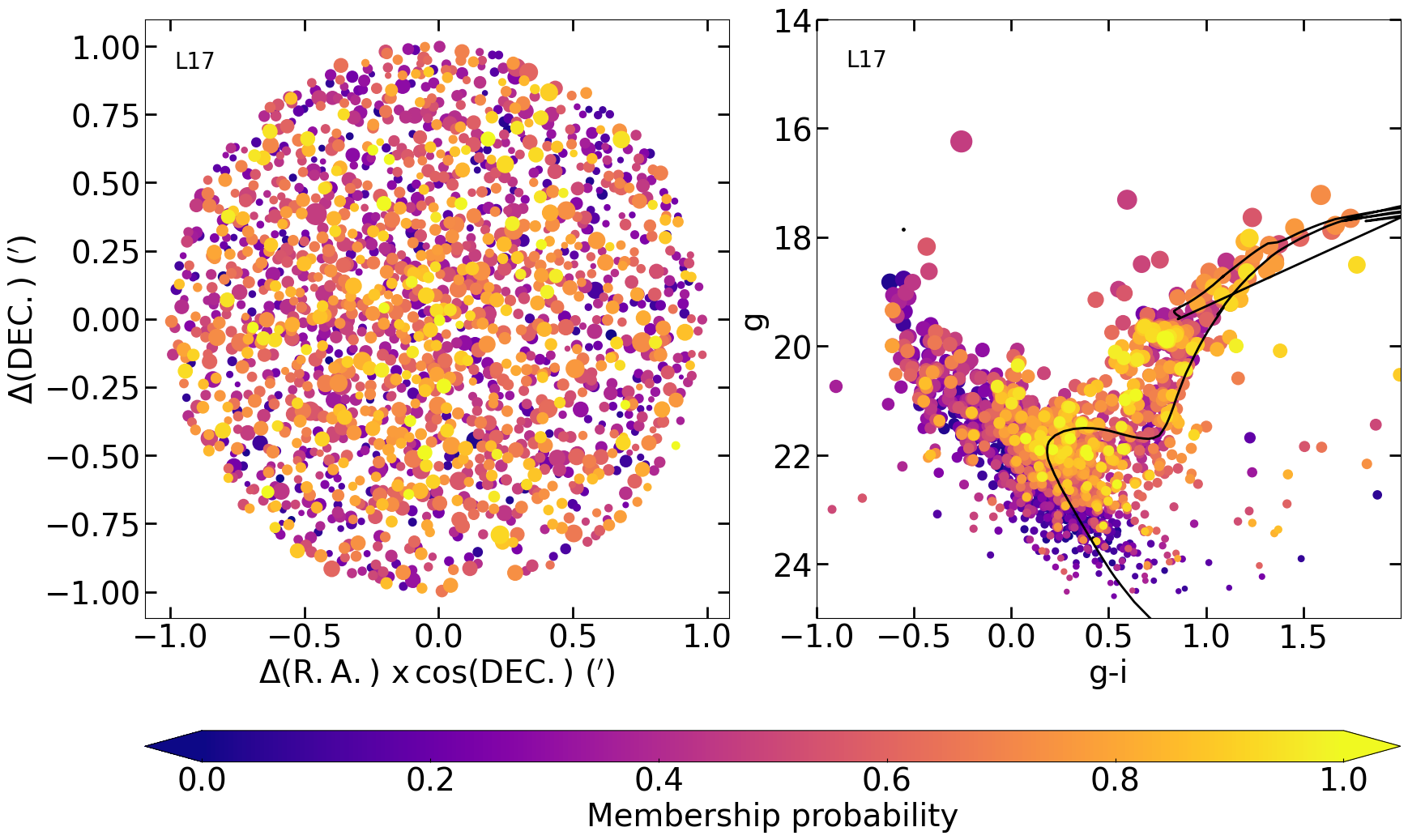}
\end{figure}
\begin{figure}[H]
\includegraphics[width=\columnwidth]{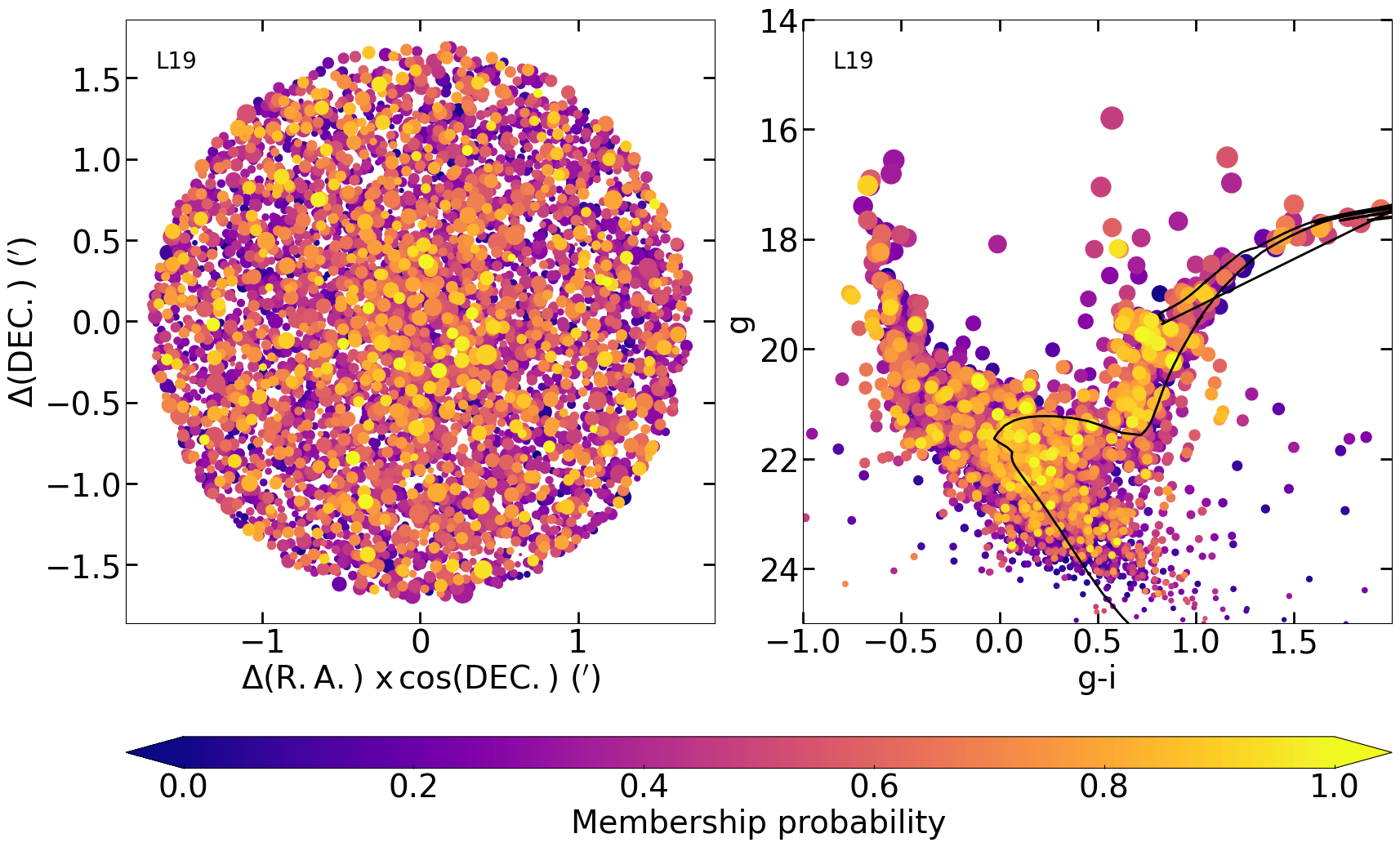}
\end{figure}
\begin{figure}[H]
\includegraphics[width=\columnwidth]{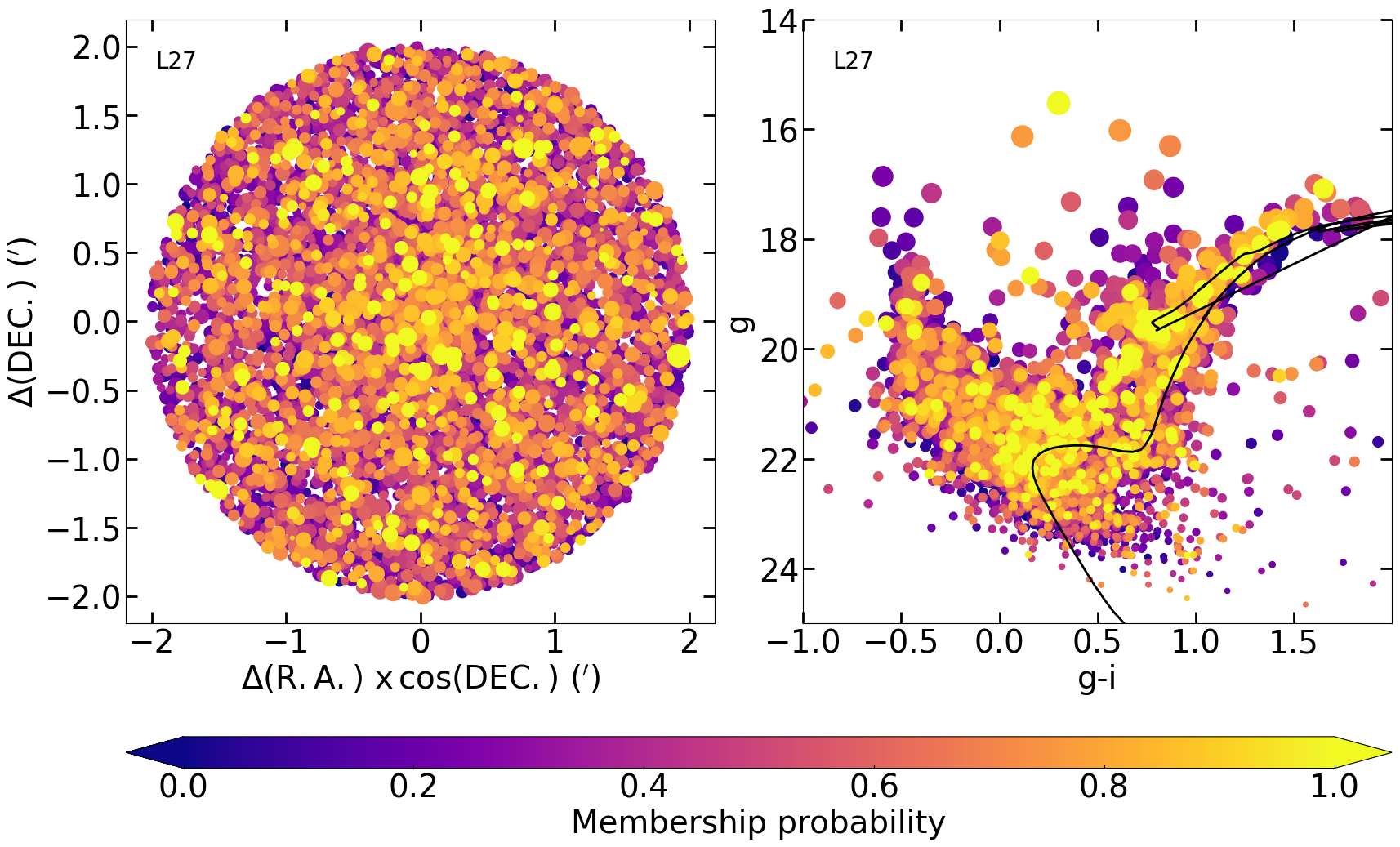}
\end{figure}
\begin{figure}[H]
\includegraphics[width=\columnwidth]{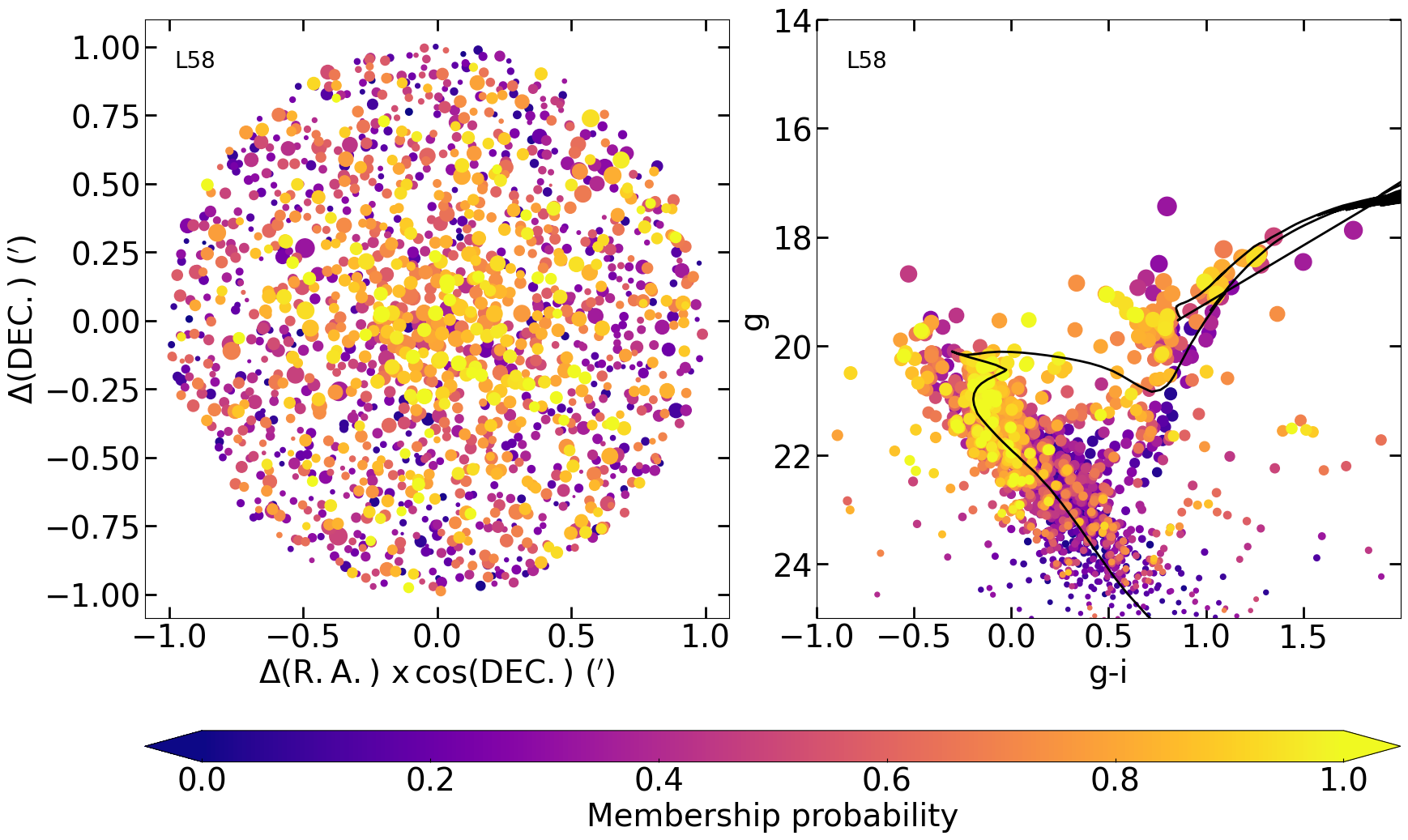}
\end{figure}
\begin{figure}[H]
\includegraphics[width=\columnwidth]{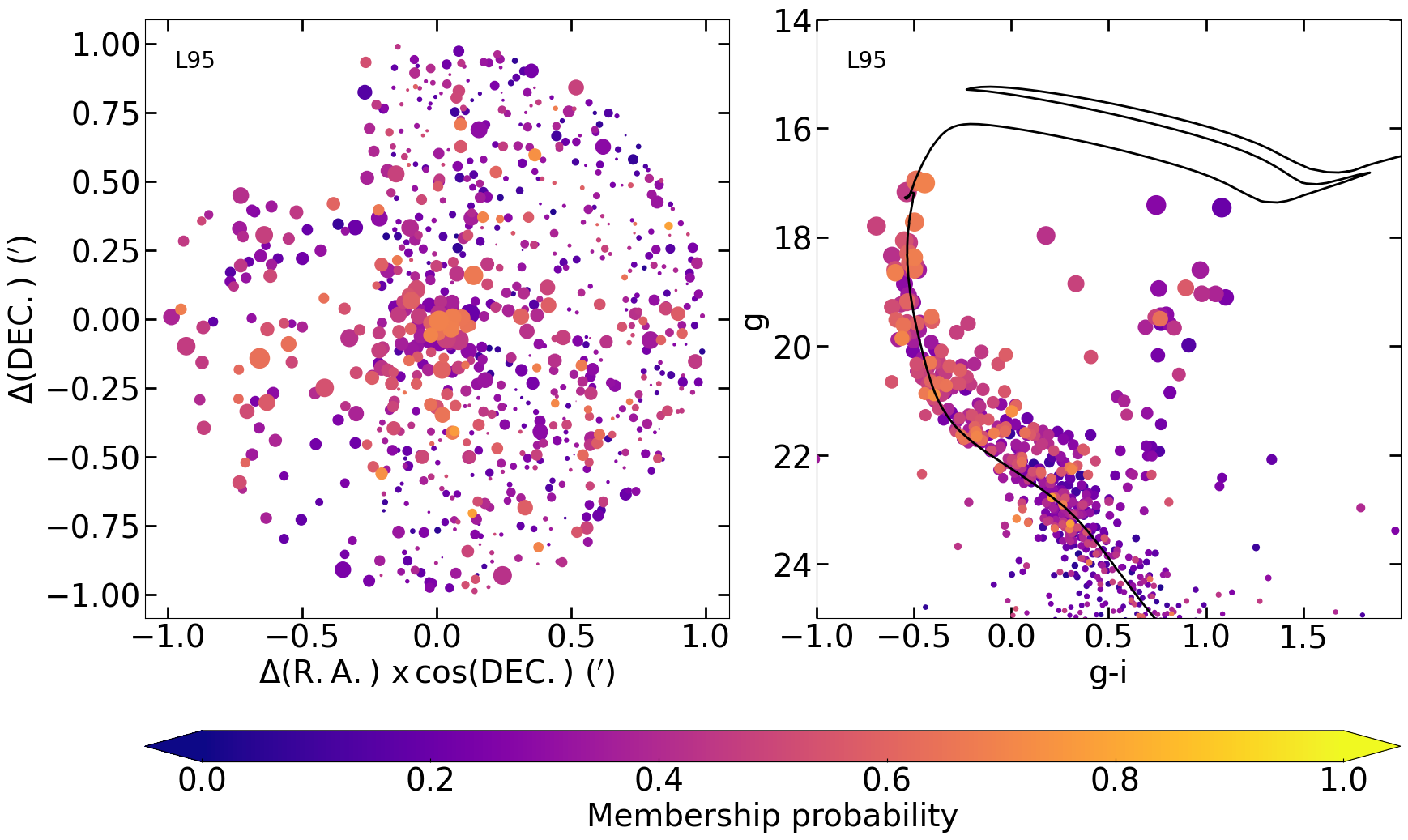}
\end{figure}
\begin{figure}[H]
\includegraphics[width=\columnwidth]{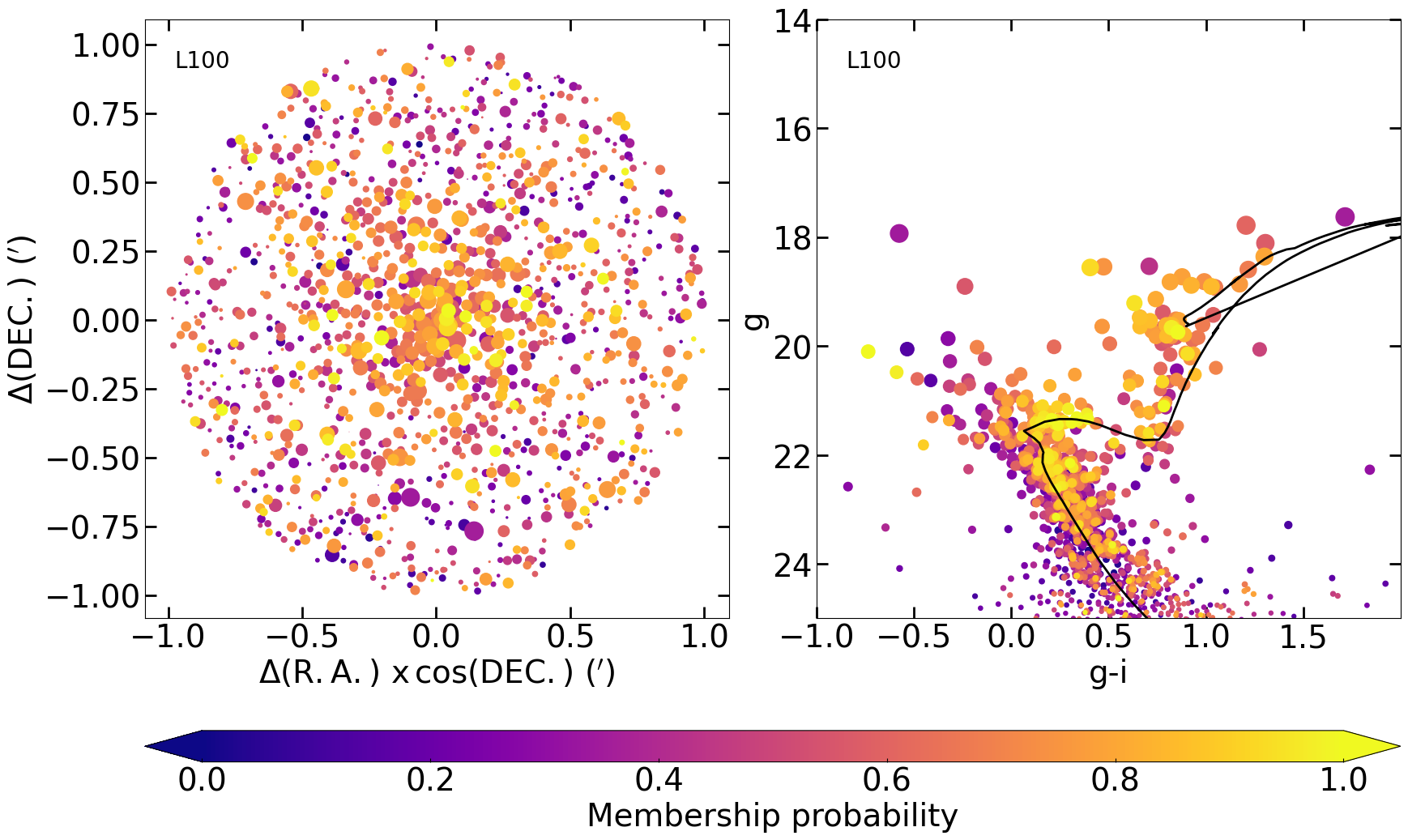}
\end{figure}
\begin{figure}[H]
\includegraphics[width=\columnwidth]{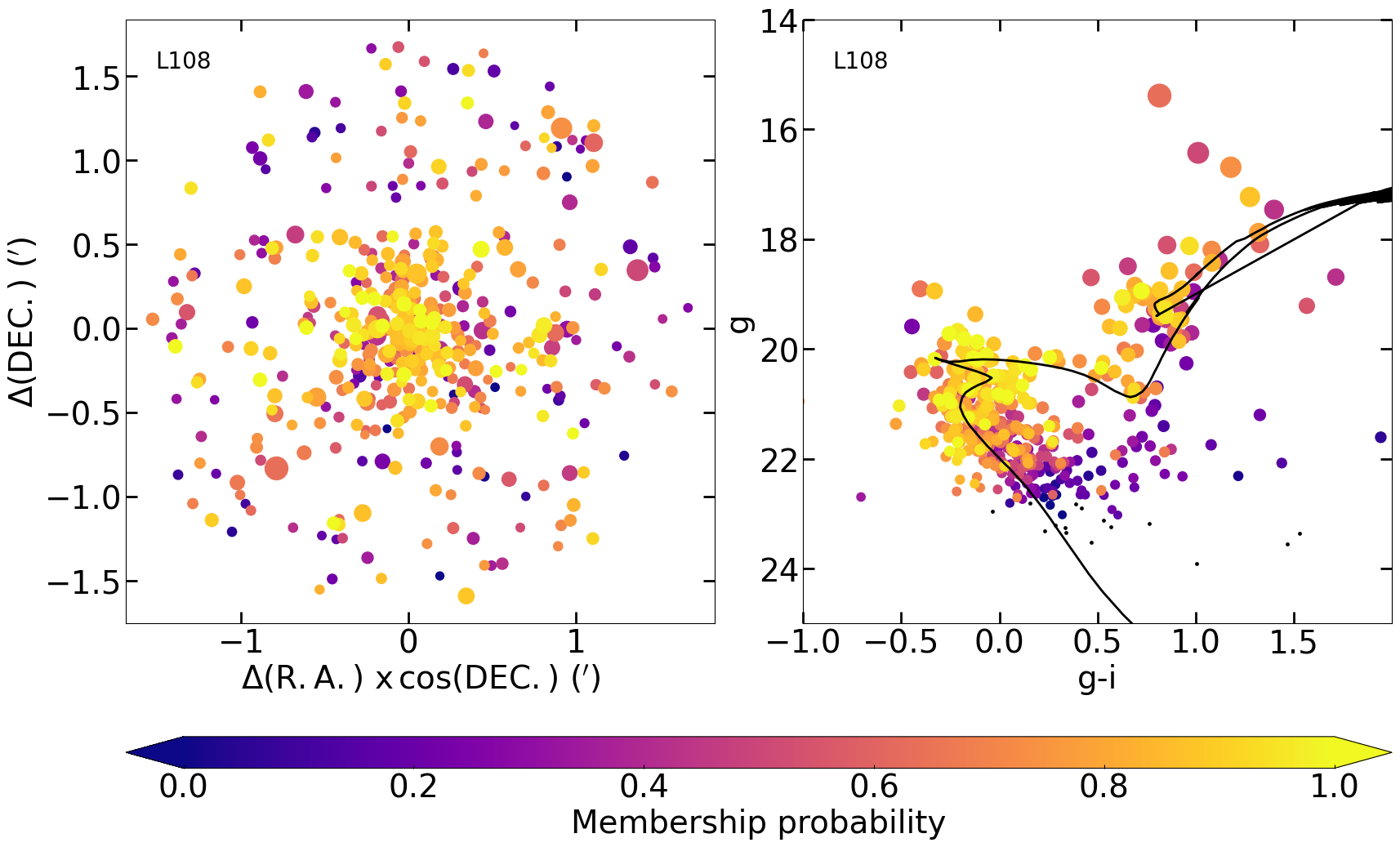}
\end{figure}
\begin{figure}[H]
\includegraphics[width=\columnwidth]{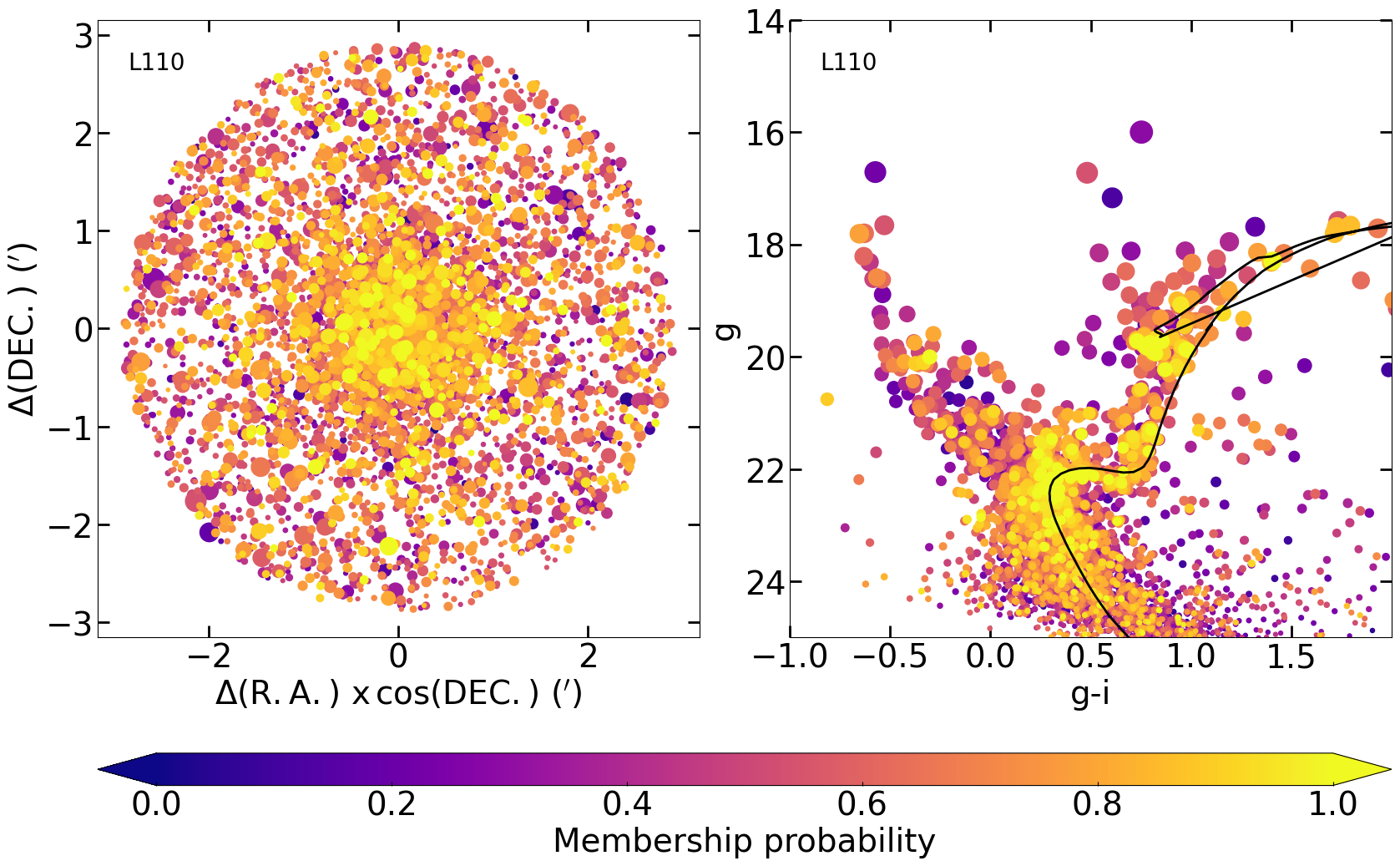}
\end{figure}
\begin{figure}[H]
\includegraphics[width=\columnwidth]{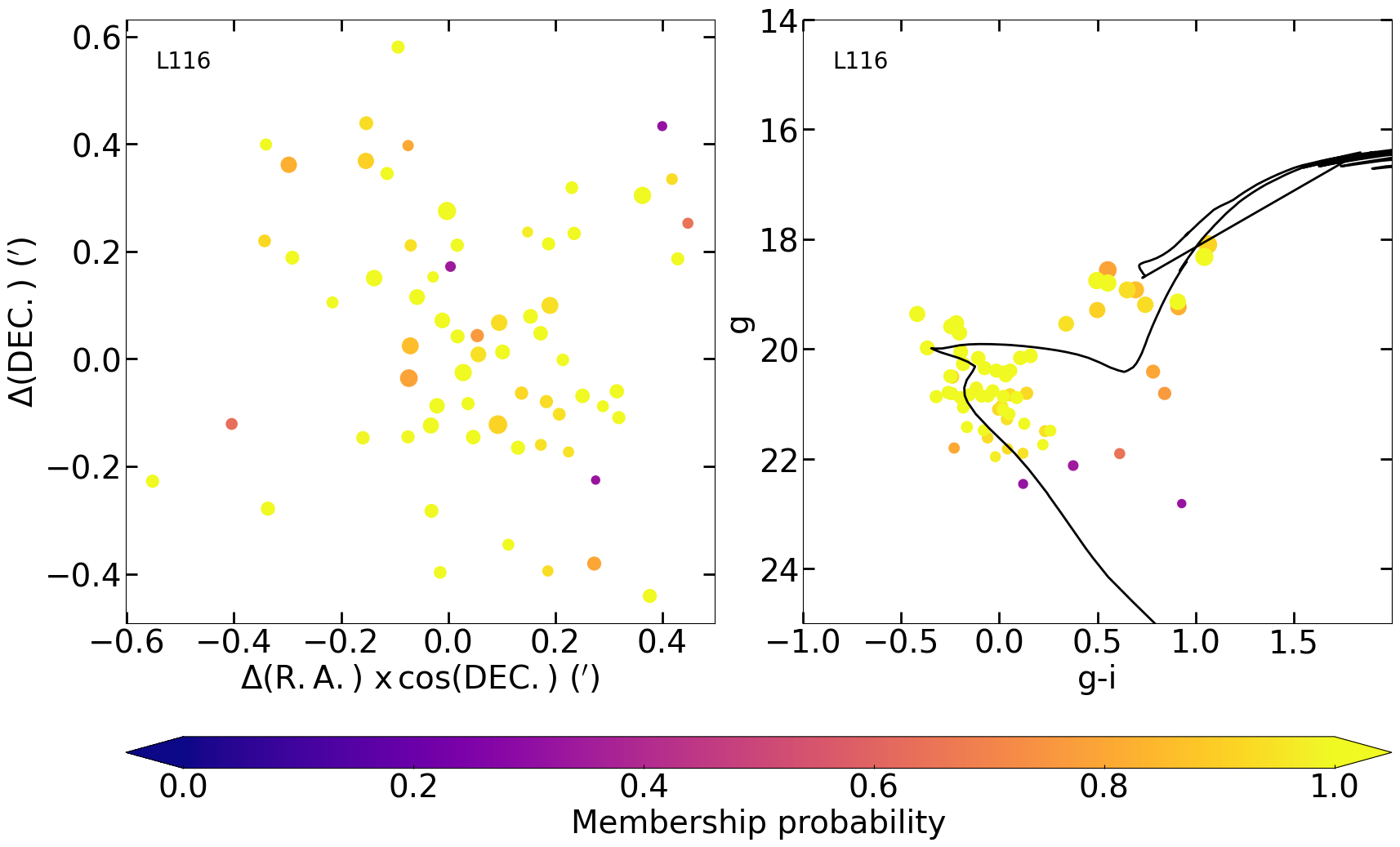}
\end{figure}
\begin{figure}[H]
\includegraphics[width=\columnwidth]{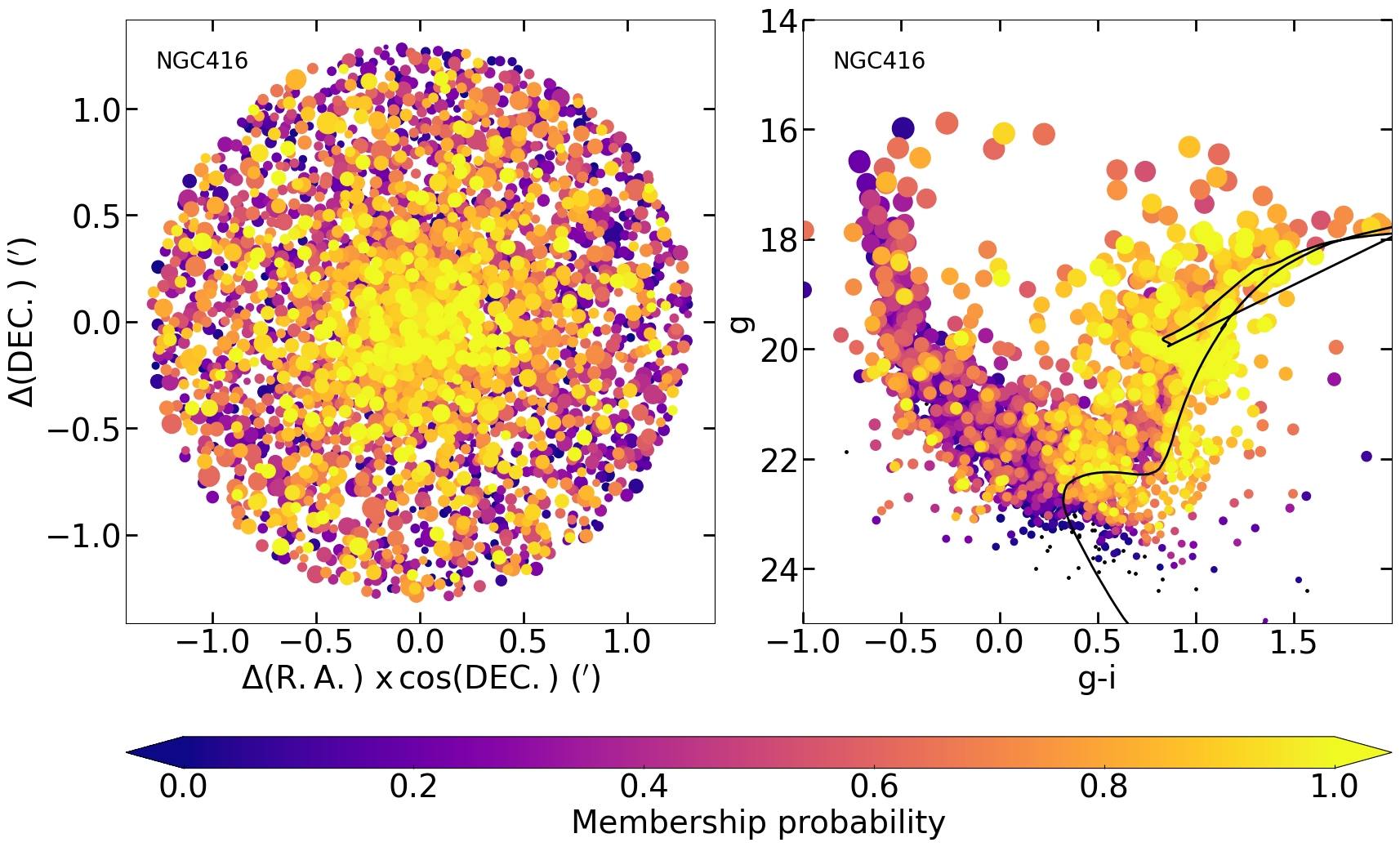}
\end{figure}
\begin{figure}[H]
\includegraphics[width=\columnwidth]{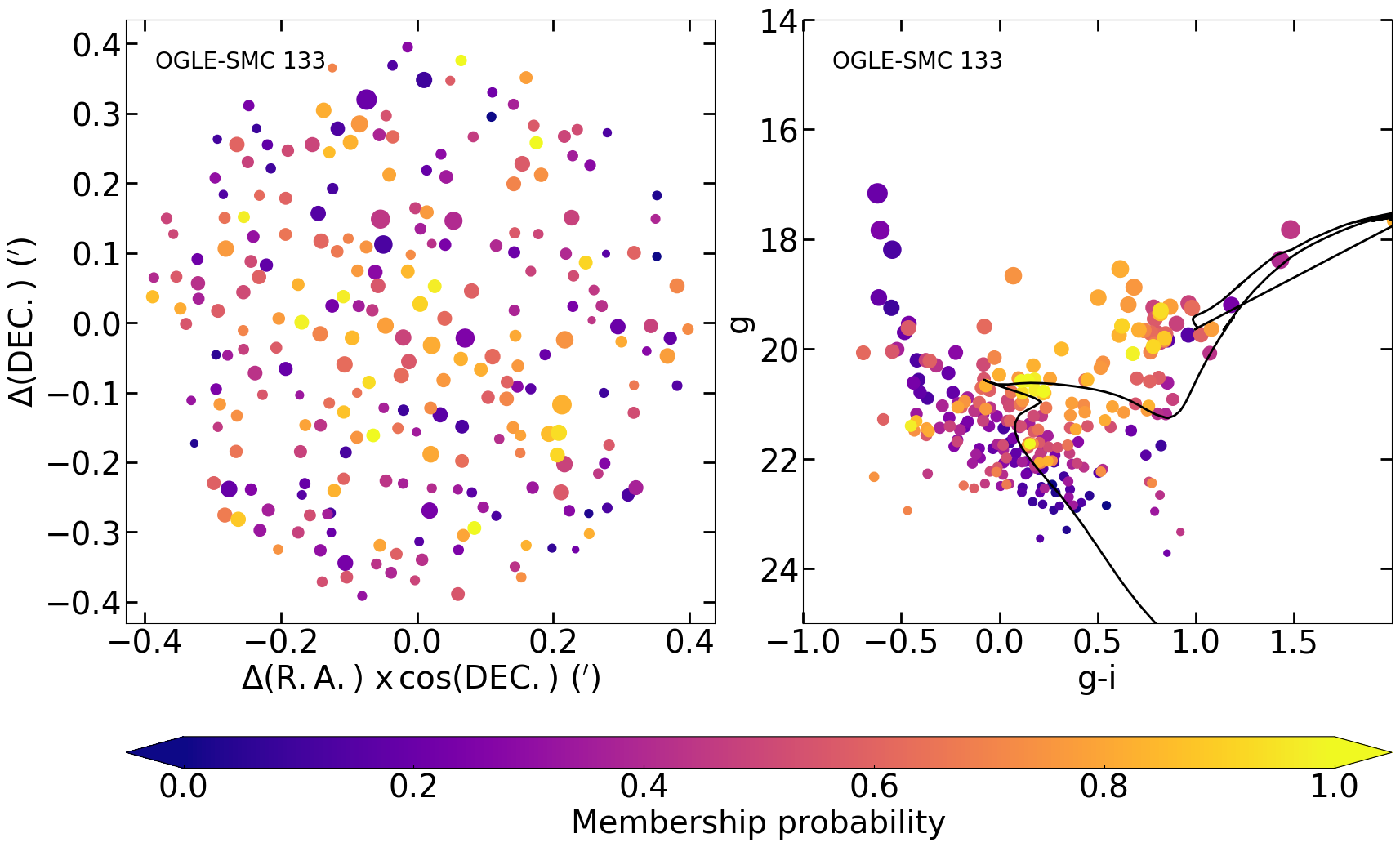}
\end{figure}

\end{document}